\documentclass[journal]{IEEEtran}

\usepackage{amsmath,amssymb,amsfonts}
\usepackage{graphicx}
\usepackage{xcolor}
\usepackage{multirow}
\usepackage{array}
\usepackage{enumitem}
\usepackage{url}

\definecolor{darkred}{rgb}{0.6,0.0,0.0}
\newcommand{\azul}[1]{\textcolor{black}{#1}}

\newcommand{\vermelho}[1]{\textcolor{black}{#1}}

\setlist[itemize]{leftmargin=10pt}
\setlist[enumerate]{leftmargin=10pt}

\begin{document}

\title{A Failure Mode Taxonomy and Severity Model for Ethereum Smart Contracts}

\author{Fernando~Richter~Vidal,
        Naghmeh~Ivaki,
        and~Nuno~Laranjeiro%
\thanks{The authors are with the University of Coimbra, CISUC/LASI, DEI, Portugal (e-mail: fernandovidal@dei.uc.pt; naghmeh@dei.uc.pt; cnl@dei.uc.pt). Corresponding author: Nuno Laranjeiro.}%
\thanks{This work is funded by national funds through FCT -- Foundation for Science and Technology, I.P., within the scope of the research unit UID/00326 -- Center for Informatics and Systems of the University of Coimbra, https://doi.org/10.54499/UID/00326/2025 and through the PhD scholarship 2023.03131.BD.}}

\markboth{IEEE Transactions on Reliability}%
{Richter Vidal \MakeLowercase{\textit{et al.}}: A Failure Mode Taxonomy and Severity Model for Ethereum Smart Contracts}

\maketitle

\begin{abstract}
Smart contracts are a core component of blockchain systems, but their immutability makes failures in deployed code catastrophic and irreversible. Existing work classifies vulnerabilities, defects, and attacks, yet the observable failures of smart contracts, the deviations from intended behavior or the expected ledger state seen once a contract is deployed, have been largely disregarded. In this paper, we design and validate a failure mode taxonomy for Ethereum smart contracts, together with a severity model that ranks failures by their observable impact. Using a literature review as input, we follow the taxonomy development method of Nickerson et al. to derive the failure modes and their dimensions, and we subject the resulting taxonomy to a saturation exercise over a systematic literature review, the OpenSCV vulnerability catalog, and smart contract incidents reported after the search cutoff. Each failure mode is characterized along five dimensions, namely root cause, violated security property, impact (recorded as primary, secondary, and other), affected component, and nature, and is given a standardized name targeting the observable malfunction rather than the underlying vulnerability. From the taxonomy, we derive CIPER, a five-level severity model that groups failure modes by their observable impact. The saturation exercise produced no new failure modes or dimensions. The taxonomy offers a shared vocabulary for describing what smart contracts do wrong when they fail, and CIPER links that vocabulary to the severity of each observable outcome.
\end{abstract}


\section{Introduction}

Smart contracts are a core component of blockchain systems~\cite{buterin2014ethereum}, executing agreements without trusted intermediaries on platforms such as Ethereum~\cite{wood2014ethereum}, and supporting token exchanges, lending and borrowing, voting, and cross-chain asset transfers~\cite{werner2022sok}. As of early 2026, the total value locked in decentralized finance protocols exceeds \$130~billion~\cite{defillama2026}, making their dependability a matter of direct economic consequence.

The immutability of blockchain deployments means a faulty contract cannot simply be patched; once deployed, its logic is permanent unless upgrade mechanisms were designed in advance~\cite{wood2014ethereum}. Failures in this domain are often catastrophic and irreversible~\cite{atzei2017survey, perez2021vulnerable}. High-profile incidents such as the DAO hack~\cite{mehar2019understanding}, the Parity wallet freeze~\cite{destefanis2018smart}, and the flash-loan exploits of the DeFi era~\cite{qin2021attacking} have shown that funds are lost permanently, protocol states become irreparably corrupted, and economic incentives are distorted across entire ecosystems.

The community has responded with a growing number of classification schemes and detection tools~\cite{vidal2024vulnerability}. Static tools such as Mythril, Slither, and Securify identify vulnerability instances at the code level, and the few examples that provide severity assessments use simple scales (e.g., critical/high/medium/low). More recent efforts include the OWASP Smart Contract Top~10~\cite{owaspsctop10_2026}, ranking based on real incident data; Iuliano and Di Nucci~\cite{iuliano2026slr}, identifying 192 vulnerabilities across 14 categories; and OpenSCV~\cite{vidal2024openscv}, covering 94 vulnerabilities mapped from 77 detection tools and research papers.

All of these efforts, however, target \textit{vulnerabilities}, that is, internal faults that allow external entities to harm the system~\cite{avizienis2004basic}. The categorization of \textit{failures}, the observable deviations from intended system behavior or the expected ledger state, has been largely disregarded in the blockchain domain. A single vulnerability may manifest as different failures depending on context, and a single failure may have multiple root causes, so a failure classification cannot really be automatically derived from a vulnerability classification. Without it, the community lacks a shared vocabulary for describing, comparing, and reasoning about how smart contracts actually fail in practice. To our knowledge, a multidimensional failure taxonomy for smart contracts, built through an established taxonomy development method, has not yet been proposed. This paper addresses the following research questions:

\begin{itemize}
  \item \textbf{RQ1:} What are the failure modes of Ethereum smart contracts?
  \item \textbf{RQ2:} How can smart contract failure modes be classified into a multidimensional taxonomy?
  \item \textbf{RQ3:} Does a failure taxonomy developed from a specific set of smart contract failures remain stable when applied to new, previously unseen failures?
  \item \textbf{RQ4:} How can Ethereum smart contract failure modes be organized into a severity-based failure model that reflects their impact on the blockchain?
\end{itemize}

\textbf{In this paper, we design and validate a failure mode taxonomy for Ethereum smart contracts}. We first conduct a literature review to identify failure modes from Ethereum smart contract incidents documented in peer-reviewed research. We then follow the taxonomy development method of Nickerson et al.~\cite{nickerson2013method} to iteratively derive classification dimensions from the failure data. Each failure mode is characterized along five dimensions: root cause, violated security property, impact (primary, secondary, and other), affected component, and nature (irreversible, transient, or permanent). A standardized name and description are assigned to each failure mode, capturing the observable malfunction rather than the underlying vulnerability. \azul{A saturation exercise with three parts, a systematic literature review carried out after the construction of the taxonomy, a mapping of the effects of every OpenSCV~\cite{vidal2024openscv} vulnerability to a failure mode, and the classification of post-cutoff incidents drawn from public trackers, produces no new failure modes or dimensions.} From the resulting taxonomy, we derive CIPER, a five-level severity model that groups failure modes by the severity of their observable impact on the blockchain.

A failure taxonomy provides value beyond classification. Smart contract incident reporting currently suffers from inconsistent, largely vulnerability-centric terminology~\cite{iuliano2026slr, vidal2024openscv, ruggiero2024sok}, making it hard to aggregate findings or compare mitigation strategies. Chillarege et al.~\cite{chillarege1992odc} showed that a structured defect taxonomy turns defect data into a quantitative measurement instrument, enabling process analysis, targeted testing, and resource allocation. A failure-oriented taxonomy for smart contracts can serve a similar purpose, giving incident analysis a shared vocabulary and highlighting which failure modes deserve the most attention.

The main contributions of this work are as follows:

\begin{itemize}
  \item A catalog of failure modes for Ethereum smart contracts, each with a standardized name and description targeting the observable failure rather than the underlying vulnerability.
  \item A multidimensional failure taxonomy that classifies failure modes along five empirically derived dimensions, built through an iterative process following Nickerson et al.~\cite{nickerson2013method}.
  \item CIPER, a five-level severity model that groups failure modes by the severity of their observable impact on the blockchain.
\end{itemize}

The iteration data, the full taxonomy, the saturation set, and the CIPER assignments are openly available as supplementary material~\cite{zenodo}.

The rest of this paper is organized as follows. Section~\ref{sec:soa} presents the background and related work. Section~\ref{Methodology} details the methodology used to build the taxonomy, including the literature review that produced the literature set. Section~\ref{sec:taxonomy} presents the failure taxonomy and the saturation exercise. Section~\ref{sec:ciper} derives the CIPER failure model. Section~\ref{sec:threats} addresses threats to validity. Finally, Section~\ref{sec:conclusion} concludes the paper.
\section{Background and Related Work}
\label{sec:soa}

This section positions our work along two axes: failure taxonomies and failure models from software and dependable systems, which provide the concepts we transfer to smart contracts; and existing classification schemes in the smart contract domain, which mostly target vulnerabilities, defects, or attacks rather than observable failures.

\subsection{Failure Modes, Taxonomies, and Failure Models}

The dependability community distinguishes faults, errors, and failures \cite{avizienis2004basic}: a fault is an internal defect, an error is the deviated internal state produced when a fault is activated, and a failure is the observable deviation of the service from its specification. Vulnerabilities are, under this view, a specific kind of internal fault; failure classifications therefore target observable behavior, not internal flaws. Avizienis et al.~\cite{avizienis2004basic} propose a multidimensional template (domain, detectability, consistency, consequences); this differs from defect classifications such as Orthogonal Defect Classification~\cite{chillarege1992odc}, which records defect attributes during inspection as an in-process measurement instrument. The CRASH severity scale of Koopman and DeVale~\cite{koopman2000ballista}, from software robustness assessment~\cite{laranjeiro2021robustness}, classifies the observable response of OS APIs under faulty inputs. Throughout this paper, we use \emph{failure model} for a condensed representation of observable failure modes derived from empirical observation~\cite{avizienis2004basic, gunawi2014cloudbugs}; CRASH is one such model.

Domain-specific characterizations have followed this pattern in cloud systems~\cite{gunawi2014cloudbugs} and deep learning. Humbatova et al.~\cite{humbatova2020dlfaults} develop a taxonomy of real deep-learning faults organized as 15 categories across three branches, validated through developer interviews and a follow-up survey. Although they classify faults rather than failures, they apply a structured taxonomy development process to an emerging domain to replace fragmented terminology with a common reference. The present paper fills the equivalent gap, on the failure side, for Ethereum smart contracts.

\subsection{Classifications in the Smart Contract Domain}

The smart contract community has produced an extensive body of classification work, but most of it targets internal faults (vulnerabilities, defects, weaknesses) or attack techniques rather than observable failures. Atzei et al. \cite{atzei2017survey} organize Ethereum attacks by the abstraction layer where the cause originates (Solidity, EVM, Blockchain). Vulnerability catalogs have since expanded: OpenSCV \cite{vidal2024openscv} consolidates 94 vulnerabilities from 77 detection tools and research papers into an open hierarchical taxonomy; Iuliano and Di Nucci \cite{iuliano2026slr} identify 192 vulnerabilities across 14 categories in an updated systematic literature review; and Ruggiero et al. \cite{ruggiero2024sok} propose a unified data model to reduce the terminological fragmentation. Closely related DeFi incident work \cite{zhou2023sokdefi} systematizes attack spaces rather than failure spaces. All these efforts share the same unit of analysis (an internal fault or an attack); none targets observable failures.

The closest previous works are Chen et al. \cite{chen2020defects} and Soud et al. \cite{soud2024fly}. Chen et al. \cite{chen2020defects} define a catalog of 20 smart contract defects extracted from 17,128 Ethereum Stack Exchange posts, associating each defect with developer-reported symptoms; the unit is a defect (an internal fault per \cite{avizienis2004basic}), and the construction relies on developer narratives rather than a documented taxonomy development method. We differ in unit (failures rather than defects), in source (incidents described in peer-reviewed research rather than developer forum posts), and in method (Nickerson-based iterative taxonomy development \cite{nickerson2013method}). Soud et al. \cite{soud2024fly} propose a classification of Ethereum code weaknesses with two parallel category sets for error source and potential impact; the explicit source/impact separation is close in intent to our separation of fault origin from observable failure, but Soud et al. \cite{soud2024fly} offer two analytical category lists while we construct a multidimensional failure taxonomy with an accompanying failure model built via a formal method.

Rezaei et al. \cite{rezaei2025rootcause} systematize 50 real-world smart contract attacks (2022--2025, over USD 1 billion in losses) and propose a four-tier root-cause framework (protocol logic design, lifecycle and governance, external dependencies, and classic smart contract vulnerabilities). Their observation that most high-impact incidents cannot be explained by individual code-level vulnerabilities aligns with ours; they focus on what originated the attack, while we focus on what the contract observably did wrong, making the two work complementary.

\subsection{Positioning Within the Literature}

Table \ref{tab:relwork} contrasts the works above along four dimensions: the \emph{unit of classification} (failure, defect, vulnerability, attack), the \emph{form of the output} (flat catalogue, multidimensional taxonomy, or condensed failure model in the sense of Koopman's CRASH), the \emph{construction method} (only some apply a named, documented process), and whether the output is \emph{multidimensional}. To the best of our knowledge, no previous smart contract work combines the four properties of the present paper: (i) observable failures as the unit of classification, (ii) a multidimensional taxonomy plus a derived failure model, (iii) the Nickerson taxonomy development method, and (iv) a literature set grounded in a literature review of documented incidents and validated through a saturation exercise.

\begin{table*}[!t]
\centering
\caption{Failure, defect, vulnerability, and attack classifications, contrasted with the present work.}
\label{tab:relwork}
\renewcommand{\arraystretch}{1.15}
\setlength{\tabcolsep}{4pt}
\begin{tabular}{p{3.2cm} p{2.6cm} p{2.4cm} p{3.2cm} p{2.8cm} p{1.4cm}}
\hline
\textbf{Work} & \textbf{Domain} & \textbf{Unit classified} & \textbf{Output} & \textbf{Method} & \textbf{Multidim.} \\
\hline
Avizienis et al. \cite{avizienis2004basic} & Dependable systems & Failure mode & Conceptual taxonomy & Conceptual & Yes \\
Koopman \& DeVale \cite{koopman2000ballista} & OS APIs (Ballista) & Failure & Failure model (CRASH) & Empirical & No \\
Chillarege et al. \cite{chillarege1992odc} & Software process & Defect & Taxonomy (ODC) & Structured & Yes \\
Gunawi et al. \cite{gunawi2014cloudbugs} & Cloud systems & Bug / failure & Empirical taxonomy & Empirical study & Yes \\
Humbatova et al. \cite{humbatova2020dlfaults} & Deep learning & Fault & Taxonomy (15 categories) & Empirical, validated & Yes \\
Atzei et al. \cite{atzei2017survey} & Smart contracts & Attack & Layered survey & Narrative & Yes (layers) \\
Chen et al. \cite{chen2020defects} & Smart contracts & Defect & Catalog (20 defects) & Empirical (forum) & No \\
Soud et al. \cite{soud2024fly} & Smart contracts & Weakness & Source + impact lists & Empirical & Partial \\
OpenSCV \cite{vidal2024openscv} & Smart contracts & Vulnerability & Hierarchical taxonomy (94) & Systematic & Yes \\
Rezaei et al. \cite{rezaei2025rootcause} & Smart contracts & Root cause & Four-tier framework & SLR of incidents & Yes \\
Zhou et al. \cite{zhou2023sokdefi} & DeFi & Attack & SoK taxonomy & SLR + incidents & Yes \\
\textbf{Present work} & \textbf{Smart contracts} & \textbf{Failure} & \textbf{Taxonomy + failure model} & \textbf{Nickerson} & \textbf{Yes} \\
\hline
\end{tabular}
\end{table*}
\section{Methodology for Building the Taxonomy}
\label{Methodology}

\subsection{Literature Set Construction}
\label{CorpusConstruction}

The literature set that feeds the iterative process was built through a literature review. We searched Google Scholar with a query combining ``smart contract'' with failure-oriented terms (``failure mode'', ``failure model'', ``failure taxonomy'', ``failure classification''), and retained peer-reviewed publications in English together with technical specifications (e.g., Ethereum Improvement Proposals) and preprints when relevant. The retained studies were screened by title, abstract, and full text for scope, retaining only failures with at least a partial origin in the smart contract, yielding the set from which we extracted failure descriptions along three analytical dimensions (root cause, affected component, and observable impact), and normalized the often vulnerability-centric terminology of the source authors into outcome-oriented labels. The full list of selected studies is available in the supplementary material~\cite{zenodo}. \azul{This construction search was a narrow, initial review; the saturation exercise (Section~\ref{SaturationExercise}) later applies a systematic literature review, with a documented and reproducible search protocol, as a post-construction check.}

\subsection{Taxonomy Development Method}
\label{TaxonomyDevelopmentMethod}

\begin{figure}[h]
\centering
	\begin{center}	\includegraphics[scale=0.35]{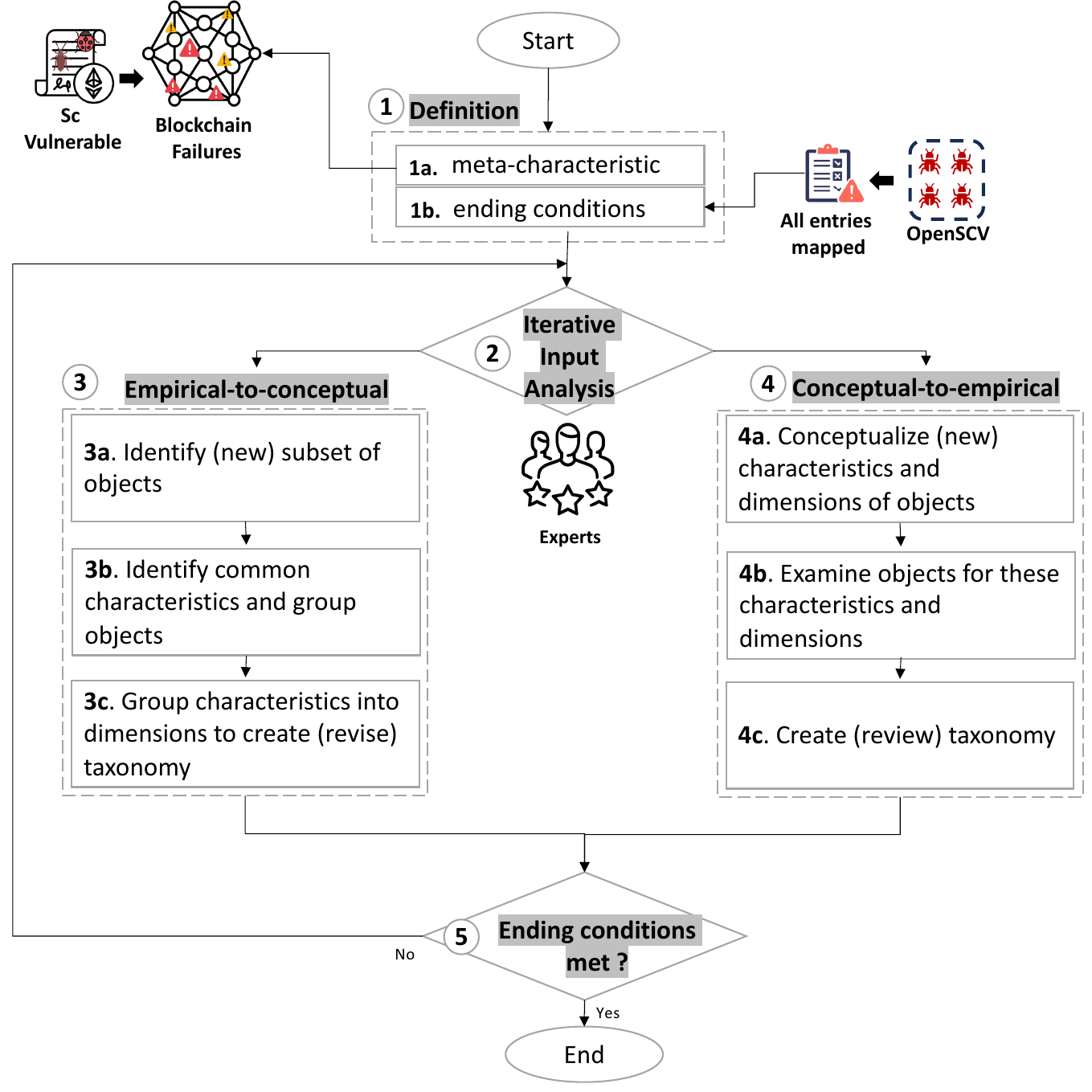}
		\caption{Taxonomy construction process (Nickerson et al.~\cite{nickerson2013method}).}
		\label{fig:taxonomydevelopment}
	\end{center}
\end{figure}

We follow the method of Nickerson et al.~\cite{nickerson2013method}, illustrated in Figure~\ref{fig:taxonomydevelopment}. It starts by defining a meta-characteristic (1a) and ending conditions (1b), and then enters an iterative cycle (2) that alternates an inductive \textbf{Empirical-to-Conceptual} step (3), in which data from the source papers informs conceptual labels, and a deductive \textbf{Conceptual-to-Empirical} step (4), in which theoretical dimensions derived from domain knowledge are validated against the empirical objects. The Conceptual-to-Empirical step also consolidates terminology, determines failure nature, identifies affected components, and associates failures with security properties. The process concludes with an evaluation step (5) confirming that the ending conditions are met.

\textbf{Meta-characteristic (1a)}: The meta-characteristic of our taxonomy is the observable impact of smart contract failures on the blockchain, capturing what a deployed contract does wrong when it fails rather than the internal defects that caused it, such as vulnerabilities, bugs, or design flaws. This outcome-oriented perspective helps practitioners and researchers assess failure severity, prioritize responses, and design mitigation strategies that reduce recurrence in future deployments. For example, given the immutable nature of blockchain systems, failures that compromise ledger integrity are particularly critical, since inconsistencies, unauthorized state changes, and irreversible corruption cannot be corrected without external intervention.

\textbf{Ending conditions (1b):} Following Kundisch et al.'s update to Nickerson's method~\cite{kundisch2022update}, we adopt three domain-appropriate ending conditions (object exhaustion, characteristic coverage, and dimensional convergence); Nickerson's other objective conditions are satisfied by construction or by stability in Iteration 5. Convergence requires:
\begin{itemize}
\item \textit{Literature exhaustion:} all papers identified during the literature search have been fully analyzed, and no new elements emerge.
\item \textit{Security Principles Coverage:} the \textit{Security Property} dimension instantiates at least the three core properties of the CIA triad (Confidentiality, Integrity, Availability)~\cite{iso27000_2018}. The dimension admits two further values, Authenticity and Non-Repudiation, for finer characterization when required.
\item \textit{Dimensional convergence:} no new dimensions or characteristics emerged in the last iteration.
\end{itemize}

\subsection{Execution of the Iterations}
\label{IterationsExecution}

Applying this methodology, the taxonomy was built through five iterations, summarised in Table \ref{tab:iterations} and described below. The three authors participated in all iterations: failure extractions, dimension definitions, and value assignments were reviewed jointly, disagreements were resolved through discussion, and the output of each iteration reflects the agreed position. The dataset produced at the end of each iteration is available in the supplementary material~\cite{zenodo}.

\begin{table}[htbp]
\centering
\caption{Summary of the five iterations.}
\label{tab:iterations}
\renewcommand{\arraystretch}{1.15}
\setlength{\tabcolsep}{4pt}
\begin{tabular}{p{0.6cm} p{2.3cm} p{4.8cm}}
\hline
\textbf{Iter} & \textbf{Method} & \textbf{Main outcome} \\
\hline
1 & E-to-C & Extraction of 23 failure modes from literature with original names; two initial dimensions (Blockchain Affected Component, Blockchain Impact). \\
2 & C-to-E & Addition of the Security Property dimension based on ISO/IEC 27000. \\
3 & C-to-E (with E-to-C refinement) & Outcome-oriented renaming of the failure modes; six-dimension structure; Blockchain Impact split into primary and secondary. \\
4 & C-to-E & Introduction of the Root Cause dimension (technical cause) and generalization of Reversibility into Failure Nature with an accompanying Reason field. \\
5 & Validation and consolidation & Dimensions Blockchain Impact, Failure Origin and Threat Nature replaced by a hierarchical Primary / Secondary / Other Impact; auxiliary fields added for traceability. \\
\hline
\end{tabular}
\end{table}

\textbf{Iteration 1 (Empirical-to-Conceptual).} Starting from the literature set (Section~\ref{CorpusConstruction}), the authors extracted 23 failure modes using the terminology employed in the original sources (e.g., \textit{Oracle trustless}, \textit{Frozen-Ether}, \textit{Out-of-Gas}). Two dimensions emerged directly from the reported data: \textit{Blockchain Affected Component} and \textit{Blockchain Impact}. The output of this iteration is the initial set of 23 failures and the two dimensions on which subsequent iterations build.

\textbf{Iteration 2 (Conceptual-to-Empirical).} The security properties defined by ISO/IEC 27000~\cite{iso27000_2018} (Confidentiality, Integrity, Availability, Authenticity, and Non-Repudiation) were applied to the set of failures as a theoretical dimension. Each failure was mapped to the property it violates, resulting in the addition of \textit{Security Property} as a new dimension. No changes were introduced to the failure set itself.

\textbf{Iteration 3 (Conceptual-to-Empirical, with empirical refinement).} The failure modes were renamed to capture the observable deviation rather than the underlying vulnerability (e.g., \textit{Oracle trustless} $\to$ \textit{Incorrect External Data Ingestion}; \textit{Frozen-Ether} $\to$ \textit{Permanent Asset Lock}; \textit{Out-of-Gas} $\to$ \textit{Premature Execution Halt}). A description field was added, and the taxonomy was structured around six dimensions: \textit{Affected Component}, \textit{Security Property}, \textit{Blockchain Impact}, \textit{Failure Origin}, \textit{Threat Nature}, and \textit{Reversibility}. Empirical inspection motivated splitting \textit{Blockchain Impact} into primary and secondary, and reclassifying some \textit{Failure Origin} values (\textit{Double-Spending} moved from Environmental Dependency to Interaction Fault).

\textbf{Iteration 4 (Conceptual-to-Empirical).} Drawing on the fault-error-failure chain of Avizienis et al.~\cite{avizienis2004basic}, \textit{Root Cause} was introduced to record the technical cause behind each failure, kept distinct from its observable manifestation. \textit{Reversibility} was generalised to \textit{Failure Nature} with values \textit{irreversible}, \textit{transient} and \textit{permanent}, accompanied by a \textit{Failure Nature (Reason)} field.

\textbf{Iteration 5 (Validation and consolidation).} \textit{Blockchain Impact}, \textit{Failure Origin}, and \textit{Threat Nature} were replaced by a hierarchical impact dimension with three fields (\textit{Primary Impact}, \textit{Secondary Impact}, \textit{Other Impacts}). Two auxiliary fields were added for traceability: \textit{References} (cited in the failure description) and \textit{Original Name} (the term used in the reference providing the most complete characterization). After this iteration, no further dimensions or characteristics emerged, closing the iterative phase.

\textbf{Evaluation (5).} At the end of Iteration 5, all three ending conditions were satisfied: \textit{Literature exhaustion}, as all papers of the literature set had been analysed without yielding new failures; \textit{Security Principles Coverage}, as the \textit{Security Property} dimension instantiates the CIA triad, with Authenticity also used and Non-Repudiation admissible but not instantiated; and \textit{Dimensional convergence}, as no new dimensions or characteristics emerged in Iteration 5. Beyond convergence, we subject the taxonomy to the saturation exercise described in Section~\ref{SaturationExercise}, following Kundisch et al.'s guidance on post-construction evaluation.

\section{Failure Mode Taxonomy}
\label{sec:taxonomy}

As a result of the iterative process described in the previous section, we obtain the taxonomy of failures. Each failure mode is characterized along a set of taxonomy dimensions and complemented by a set of auxiliary fields kept for traceability and tabular presentation. Table \ref{tab:dims} lists both groups.

\begin{table}[htbp]
\centering
\caption{Taxonomy dimensions and auxiliary characteristics.}
\label{tab:dims}
\renewcommand{\arraystretch}{1.15}
\setlength{\tabcolsep}{4pt}
\begin{tabular}{p{2.6cm} p{5.4cm}}
\hline
\multicolumn{2}{l}{\textit{Taxonomy dimensions}} \\
\hline
Root Cause & Underlying technical reason that leads to the failure (e.g., logic errors, improper validation, external dependency issues). \\
Security Property & Security property violated; admissible values are the five of ISO/IEC 27000 (confidentiality, integrity, availability, authenticity, non-repudiation). \\
Impact & Consequence of the failure, recorded hierarchically at three levels: primary (dominant), secondary (follow-on), and other (indirect or less significant). \\
Affected Component & Part of the system that is impacted (e.g., contract state, ledger, accounts, network resources). \\
Failure Nature & Persistence characteristic (irreversible, permanent, transient). \\
\hline
\multicolumn{2}{l}{\textit{Auxiliary characteristics}} \\
\hline
Description & Concise explanation of how the failure manifests. \\
Failure Nature (Reason) & Why the failure is irreversible, permanent, or transient. \\
OpenSCV Classification & Associated entries of the OpenSCV vulnerability taxonomy, when applicable; populated during the saturation exercise (Section \ref{SaturationExercise}), not during the iterative process. \\
Original Name & Term used in the reference chosen to source the failure name. \\
References & References cited in the failure description. \\
\hline
\end{tabular}
\end{table}

In the following subsections, we first describe each failure mode according to the above dimensions and then present how the saturation conditions are met.

\subsection{Failure Modes}

Each failure mode is described below along the dimensions of Table~\ref{tab:dims}.

\subsubsection{\textbf{Double-Spending}}
This failure manifests as multiple successful transactions spending the same units, originating in the consensus and smart contract layers due to validation or agreement failures. The root cause lies in inconsistent state perception and finalization under asynchronous or adversarial conditions.
Cross-chain interoperability and Layer-2 architectures~\cite{mukherjee2025doublespending,sun2024doubleup} have reintroduced this risk, mitigated in standalone blockchains. Attackers use, e.g., eclipse attacks or transaction delay manipulation to bypass consensus and spend the same assets across bridging protocols.
\textit{Integrity} is violated, leading to inconsistent ledgers, incorrect balances, financial loss, and loss of trust. The affected component is the global state (ledger and balances). The failure is \textit{irreversible}, as finalized conflicts are permanently recorded on-chain and cannot be corrected without external intervention such as forks or governance.
Commonly referred to as ``Double-Spending''~\cite{mukherjee2025doublespending}.

\subsubsection{\textbf{Incorrect External Data Ingestion}}
This failure manifests when smart contracts operate on incorrect off-chain data, leading to erroneous state transitions. It originates in the smart contract, in its unsafe consumption of external data without validating correctness or authenticity; the external oracle source supplies the faulty data, but the failure is realized when the contract acts on it without safeguards. Reliance on external entities breaks the trustless model and introduces centralization and single points of failure~\cite{hassan2023trust}; authentication flaws~\cite{hassan2023trust} or price manipulation~\cite{deng2024safeguarding} at the source propagate into the on-chain state.
\textit{Integrity} is violated, leading to incorrect state updates, financial loss, manipulation risk, and loss of trust. The affected component is the contract state (storage and ledger values dependent on external data). The failure is \textit{irreversible}, as erroneous inputs, once committed, are permanently recorded on-chain and cannot be retroactively corrected without external intervention.
Commonly referred to as ``Oracle trustless''~\cite{hassan2023trust}.

\subsubsection{\textbf{Transaction History Disclosure}}
This failure manifests when on-chain transaction activity becomes linkable to real-world identities, exposing user behavior. It originates in the smart contract, in the identity-linkable data it stores, emits, or exposes on-chain; ledger transparency is the enabling condition rather than the origin, as public ledgers permanently record the activity~\cite{buterin2024privacy}. Identity-mapping services such as \textit{Ethereum Name Service}, themselves deployed as smart contracts, exacerbate the problem, and privacy tools (e.g., mixers) fail to mitigate it because attackers apply on-chain heuristics to deanonymize and correlate addresses.
\textit{Confidentiality} is violated, leading to loss of privacy, exposure of financial behavior, profiling risks, and loss of trust. The affected component is the account and identity space (addresses and transaction histories). The failure is \textit{permanent} because once disclosed, on-chain data cannot be erased, and privacy cannot be restored even if identities are later obfuscated.
Commonly referred to as ``Lack Privacy on Transaction History''~\cite{buterin2024privacy}.

\subsubsection{\textbf{Unintended State Visibility}}
This failure manifests when the contract state intended to be private becomes publicly accessible on-chain. It originates in the smart contract layer, where the root cause lies in design misconceptions and improper handling of sensitive data, combined with incorrect visibility settings. Blockchain transparency makes all contract data globally readable~\cite{wu2024comprehensive}; simple flaws such as missing visibility modifiers (e.g., SWC-100) expose restricted data, enabling adversaries to monitor and extract confidential information.
\textit{Confidentiality} is violated, leading to exposure of sensitive data, privacy loss, exploitation risks, and loss of trust. The affected component is the contract state (storage variables and on-chain data). The failure is \textit{permanent}, as blockchain transparency makes exposed data permanently visible once deployed, and it cannot be hidden or removed.
Commonly referred to as ``Secret Field Value''~\cite{wu2024comprehensive}.

\subsubsection{\textbf{Premature Execution Halt}}
This failure occurs when transaction execution stops before completing its intended operations; the most common trigger is exceeding the gas limit~\cite{ma2023vgas}, though the same behavior can arise from other execution-halt conditions. It originates in the smart contract layer, where the root cause lies in improper gas management and inefficient execution logic, often due to compiler underestimation or adversarial inputs triggering excessive computation. Attackers may use unbounded loops or expensive operations that exceed block gas capacity, making it a vector for denial-of-service attacks.
\textit{Integrity} is violated, as intended operations fail to complete, disrupting execution flow. The impact includes failed workflows, incomplete operations, disruption of dependent actions, and reduced reliability. The affected component is the execution state, specifically transaction outcomes and their effects on system state. The failure is \textit{transient}, as it is execution-dependent, and the affected transaction can be retried or recovered in subsequent executions after gas adjustment or code optimization.
Commonly referred to as ``Out-of-Gas''~\cite{ma2023vgas}.

\subsubsection{\textbf{Silent Operation Abort}}
This failure~\cite{hajdu2020using} occurs when a transaction silently terminates upon an invocation error, halting execution without completing operations, returning values, or updating blockchain state. It originates in the smart contract layer, where the root cause lies in improper error handling and unchecked execution failures due to the misuse of abort mechanisms or logical flaws that suppress internal exceptions. The failure remains hidden from users with no explicit error message; its sole observable manifestation is the silent and unrecoverable depletion of transaction gas.
\textit{Integrity} is violated, as execution results become ambiguous and do not reflect intended logic. The impact includes inconsistent or missing state updates, ambiguous outcomes, incorrect assumptions by dependent contracts or users, and reduced diagnosability and reliability. The affected component is the contract execution state, particularly the visibility and correctness of execution outcomes. The failure is \textit{transient}, as it affects individual executions without permanently corrupting system state, and correct behavior resumes in subsequent executions.
Commonly referred to as ``Abort Failure''~\cite{hajdu2020using}.

\subsubsection{\textbf{Inconsistent State Transition}}
This failure occurs when a smart contract executes successfully but produces an unintended or incorrect state transition~\cite{bose2022sailfish}. It originates in the smart contract layer; the root cause lies in logic or specification errors that violate intended state invariants. Because execution does not revert, logical errors are committed directly to the blockchain state, making the corruption persistent and requiring external intervention or contract redeployment~\cite{vidal2025analyzing}.
\textit{Integrity} is violated, as the system evolves into states that do not match expected behavior. The impact includes invalid state, incorrect outputs, invariant violations, and cascading errors in dependent contracts or computations. The affected component is the contract state, particularly storage variables and system-level state representations. The failure is \textit{irreversible}, as committed state transitions become part of the immutable ledger and cannot be corrected without external mechanisms such as forks or redeployment.
Commonly referred to as ``Reliability Failure''~\cite{hajdu2020using}.

\subsubsection{\textbf{Corrupted Value Computation}}
This failure occurs when a computation produces a numerically incorrect result that propagates into the contract state or ledger; typical triggers are arithmetic overflows or underflows executing without runtime exceptions~\cite{vidal2025bbench,torres2018osiris}, though the same behavior can arise from other computation or type-handling errors. Because the EVM lacks native safeguards for such errors, these flawed operations bypass validation and are permanently recorded on-chain, potentially leading to large-scale financial and transactional distortions. It originates in the smart contract layer, where insufficient validation or incorrect type handling produces incorrect numerical results.
\textit{Integrity} is violated, as invalid numerical results are treated as correct and propagated through the system state. The impact includes incorrect state values, propagation of erroneous computations, financial inconsistencies, and cascading computational errors affecting dependent contracts or logic. The affected component is the contract state and ledger values, particularly numeric variables such as balances, prices, and financial data. The failure is \textit{irreversible}, as incorrect values are permanently stored on-chain and may further propagate into future computations, requiring external remediation such as corrective transactions or contract upgrades.
Commonly referred to as ``Integrity Failure''~\cite{hajdu2020using}.

\subsubsection{\textbf{Stale State Operation}}
This failure occurs when a transaction completes successfully and returns an expected result while silently operating on outdated or inconsistent state, corrupting the blockchain's read/write sets without triggering an abort~\cite{hajdu2020using}. Although the user receives a correct response, the state divergence remains hidden and may trigger cascading anomalies such as redundant transactions and increased gas consumption. It originates in the smart contract layer; the root cause lies in concurrency-related design flaws such as stale reads, improper synchronization of state updates, and incorrect assumptions about transaction ordering.
\textit{Integrity} is violated, as correct outputs are produced while the underlying state evolution is incorrect. The impact includes incorrect results due to outdated data, inconsistent state updates, concurrency anomalies, and silent logical failures that are difficult to detect. The affected component is the contract storage and system state, particularly read/write sets and execution-time state snapshots. The failure is \textit{transient}, as its effects depend on execution timing and ordering and can often be corrected through subsequent state resynchronization.
Commonly referred to as ``Latent Integrity''~\cite{hajdu2020using}.

\subsubsection{\textbf{Wasted Execution}}
This failure occurs when a transaction consumes gas executing operations that produce no useful change in state, wasting resources without effect on contract outcomes~\cite{pacheco2023ethereum,ma2023vgas}. Typical triggers include redundant computations, dead code paths, unreachable branches, or poorly optimized logic kept in the compiled contract. It originates in the smart contract layer and is caused by inefficient execution design and gas mismanagement (unnecessary computations, poorly optimized logic, incorrect estimation of resource requirements).
\textit{Availability} is violated, as contract functionality becomes economically or computationally inaccessible despite the system remaining operational. The impact includes unnecessary gas consumption, higher transaction costs, reduced efficiency, and user discouragement due to unreliable execution. The affected component is the execution layer, affecting gas consumption and computational throughput. The failure is \textit{transient}, as it is execution-dependent, produces only resource inefficiency rather than permanent state damage, and can be mitigated through code optimization.
Commonly referred to as ``Ineffective''~\cite{hajdu2020using}.

\subsubsection{\textbf{Involuntary Contract Termination}}
This failure occurs when a smart contract is prematurely destroyed or disabled by unauthorized or improperly governed actions, causing loss of functionality and state~\cite{khan2023dynamic,nikolic2018finding}. Although mitigations such as EIP-6780~\cite{eip6780} have altered the behavior of contract destruction, the underlying risk persists through application-layer termination logic and forced withdrawal mechanisms. It originates in the smart contract layer and is caused by design flaws and insufficient access control over privileged or destructive operations.
\textit{Availability} is violated, as the contract becomes inaccessible or permanently disabled for legitimate users. The impact includes irreversible loss of contract functionality, deletion or disabling of code and state, loss of user assets, and disruption of dependent systems. The affected component is the deployed contract instance, including both executable code and stored state. The failure is \textit{irreversible}, as destroyed or disabled contracts cannot be restored within the system and require external redeployment or intervention.
Commonly referred to as ``Self-Destruct''~\cite{khan2023dynamic}.

\subsubsection{\textbf{Permanent Asset Lock}}
This failure occurs when digital assets become permanently inaccessible within a smart contract or address: the ledger continues to record the balance correctly, but no contract function can release or transfer it~\cite{atzei2017survey}. The condition is often identified by static analysis tools such as MAIAN~\cite{nikolic2018finding}, which detect logic flaws that trap funds indefinitely. It originates in the smart contract layer and is caused by design, logic, or access control flaws that prevent legitimate withdrawal or recovery of assets.
\textit{Availability} is violated, as assets become inaccessible to all parties, including legitimate users. The impact includes permanent inaccessibility of funds, inability to withdraw or recover assets, financial damage, and long-term loss of trust. The affected component is the contract storage and balance layer, particularly token holdings and withdrawal-related state variables. The failure is \textit{irreversible}, as locked assets cannot be recovered through normal execution and may require external intervention, if recovery is possible at all.
Commonly referred to as ``Frozen-Ether''~\cite{wu2024comprehensive}.

\subsubsection{\textbf{Persistent Unavailability}}
This failure occurs when a smart contract cannot provide its intended service because legitimate users cannot interact with it, or its functions remain perpetually blocked. Typical triggers include malicious or failing external calls that revert and block subsequent execution (e.g., within fund-transfer loops)~\cite{wu2024comprehensive}, and design flaws that lead to gas exhaustion and prevent the function from completing. It originates in the smart contract layer and is caused by flaws in execution flow control, unsafe handling of external calls, or insufficient isolation of user interactions, where a single failing or malicious participant can block the entire execution pipeline.
\textit{Availability} is violated, as legitimate users cannot interact with the contract. The impact includes failed or blocked transactions, loss of service availability, disruption of dependent systems, reduced reliability, and loss of trust. The affected component is the service layer, specifically the contract interface and its callable functions. The failure is \textit{permanent}, as the immutability of smart contracts makes it impossible to restore the affected contract to normal operation without external intervention, typically redeployment of a fixed version.
Commonly referred to as ``Denial of Service''~\cite{wu2024comprehensive}.

\subsubsection{\textbf{Misdirected Token Transfer}}
This failure occurs when tokens are delivered to an unintended destination or lost entirely, with the sender unable to recover them; a frequent trigger is privileged token functions that allow unauthorized or unintended modifications of token balances~\cite{chen2019tokenscope,ma2023pied}, though similar outcomes arise from address-encoding mistakes or unchecked transfer arguments. This is enabled by overly permissive administrative logic in ERC token contracts and has been demonstrated in real-world exploits such as the SoarCoin incident~\cite{frowis2023detecting}. It originates in the smart contract layer and is caused by incorrect transfer logic combined with insufficient access control; the root cause lies in improperly implemented administrative functions that allow unintended or unauthorized modification of balances and ownership records.
\textit{Integrity} is violated, as token ownership and transfer semantics are no longer correctly enforced. The impact includes misdirected token transfers, irreversible loss of funds, incorrect ownership records, and loss of trust. The affected component is the token state, particularly balance mappings and ownership records in ERC contracts. The failure is \textit{irreversible}, as misdirected transfers cannot typically be reverted without recipient cooperation or external intervention.
Commonly referred to as ``Token Transmission Lost''~\cite{wu2024comprehensive}.

\subsubsection{\textbf{Nested Execution Collapse}}
This failure occurs when deeply nested calls or looping execution paths fail to complete due to stack limits or unbounded computation, causing silent execution collapse and incomplete state updates~\cite{chen2020survey,wu2024comprehensive}. While historically linked to the EVM call-stack limit (1,024 depth) and mitigated by EIP-150, similar issues persist through inefficient recursion or poorly bounded loops. It originates in the smart contract layer and is caused by improper handling of nested calls and execution depth (unbounded loops, inefficient recursion, missing safeguards against resource or stack exhaustion); the root cause lies in execution logic flaws that fail to anticipate worst-case paths, leading to incomplete or partially executed workflows.
\textit{Integrity} is violated, as execution produces incomplete or inconsistent state updates without explicit failure signaling. The impact includes incomplete execution, dropped dependent updates, partial workflow completion, and silent inconsistencies across contract interactions. The affected component is the contract state, particularly storage updates dependent on chained or nested operations. The failure is \textit{transient}, as it depends on execution path complexity and input conditions and can often be mitigated through bounded execution or improved design.
Commonly referred to as ``Stack Limit''~\cite{wu2024comprehensive}.

\subsubsection{\textbf{Distorted Economic Output}}
This failure occurs when a contract produces financial outcomes (prices, fees, payouts) that deviate from fair market conditions; a common trigger in oracle-dependent DeFi applications is manipulated or unrepresentative market data~\cite{xi2024pomabuster,li2023demystifying}, though similar outcomes arise from flawed internal pricing logic or incentive design. Attackers can exploit liquidity imbalances, wash trading, or transaction ordering strategies (often via infrastructure such as Flashbots) to manipulate price signals while evading detection. It originates in the smart contract layer, in economic and pricing logic that relies on external oracle inputs; the root cause lies in insecure dependence on external market data and insufficient validation against adversarial manipulation or abnormal liquidity conditions.
\textit{Integrity} is violated, as the system produces financial outputs that do not reflect true market conditions. The impact includes incorrect pricing, financial loss, manipulated payouts, exploitation of incentive mechanisms, and loss of trust. The affected component is the financial state of the contract, particularly price feeds, balances, and payout computations derived from Oracle data. The failure is \textit{irreversible}, as distorted economic outputs, once committed, propagate into finalized state changes and downstream transactions.
Commonly referred to as ``Price manipulation''~\cite{xi2024pomabuster}.

\subsubsection{\textbf{Duplicated Fund Disbursement}}
This failure occurs when a single valid authorization leads to multiple fund withdrawals due to replayable or improperly protected payment logic~\cite{kushwaha2022systematic,wang2025one}. Historically associated with reentrancy attacks, modern variants commonly arise from signature replay vulnerabilities enabled by missing nonces, timestamps, or expiration checks. It originates in the smart contract layer and is caused by incorrect handling of authorization and payment logic (insufficient replay protection, weak state tracking of previous executions); the root cause lies in verification mechanisms that fail to ensure the uniqueness and freshness of payment approvals, allowing the same authorization to be reused.
\textit{Integrity} is violated, as the system incorrectly executes multiple payouts from a single legitimate authorization. The impact includes repeated or excessive payments, depletion of contract balances, financial loss, exploitation of payment logic, and loss of trust. The affected component is the contract state governing balances and authorization records, particularly signature validation and payment execution logic. The failure is \textit{irreversible} because once funds are disbursed multiple times, they cannot really be automatically recovered without external intervention or recipient cooperation.
Commonly referred to as ``Multiple Payments''~\cite{wang2025one}.

\subsubsection{\textbf{Network Throughput Degradation}}
This failure occurs when resource exhaustion attacks and inefficient execution patterns overload the network, causing reduced transaction throughput, higher fees, and increased confirmation delays~\cite{perez2020broken,kushwaha2022systematic}. Attackers exploit weaknesses in EVM resource metering or gas pricing, executing computationally heavy operations at disproportionately low cost and driving system-wide congestion. It originates in the smart contract, in uncontrolled resource consumption by adversarial or inefficient contract execution that exploits under-priced EVM operations; the resulting system-wide congestion is where the impact is felt, at the network level, rather than where the failure originates. The root cause lies in excessive computational and bandwidth consumption driven by such execution, combined with insufficient cost enforcement in gas pricing that fails to reflect resource usage.
\textit{Availability} is violated, as the system's ability to process transactions efficiently is degraded for all participants. The impact includes slower confirmations, increased fees, reduced throughput, degraded user experience across the ecosystem, and loss of trust. The affected component is the network and consensus infrastructure, particularly transaction propagation and block processing capacity. The failure is \textit{transient}, as performance typically recovers once congestion or attack pressure decreases and the network capacity returns to normal.
Commonly referred to as ``Uncontrolled Resource Consumption''~\cite{kushwaha2022systematic}.

\subsubsection{\textbf{Unauthorized Action Execution}}
This failure occurs when privileged functions are executed by unauthorized entities, either because access control is broken or insufficient~\cite{kushwaha2022systematic}, or because the authentication that gates those functions is defeated, for instance through a forged or compromised signature. In Solidity, this often arises from misconfigured validation modifiers or misuse of authentication variables such as \texttt{tx.origin}, enabling attackers to bypass authorization checks via intermediary contracts. It originates in the smart contract layer and is caused by access control and authentication flaws that allow privileged operations without proper identity verification; the root cause lies in incorrect or insufficient authorization logic (missing access modifiers, weak role-based controls, improper caller validation), or in a compromised authentication primitive such as signature forgery.
\textit{Authenticity} is violated, as the system cannot guarantee that only legitimate actors execute privileged actions. The impact includes unauthorized function execution, malicious or unintended state changes, financial loss, and loss of trust. The affected component is the contract state and permission system, particularly the access control and authentication logic for sensitive functions such as transfers and administrative operations. The failure is \textit{irreversible}, as unauthorized actions directly modify persistent on-chain state and cannot really be automatically undone.
Commonly referred to as ``Insufficient Authorization''~\cite{kushwaha2022systematic}.

\subsubsection{\textbf{Unchecked Input Acceptance}}
This failure occurs when a smart contract fails to validate input parameters, often due to missing or incorrectly implemented \texttt{require} statements~\cite{munir2023statically}. Malicious or malformed inputs bypass execution constraints, potentially enabling unauthorized fund transfers and incorrect contract behavior, especially in cross-contract interactions. It originates in the smart contract layer and is caused by insufficient input validation and missing enforcement of preconditions on function arguments; the root cause lies in inadequate checks on input correctness (missing bounds, format, or authorization validation before state-modifying execution).
\textit{Integrity} is violated, as unvalidated inputs lead to unintended and incorrect state transitions. The impacts include processing of invalid inputs, unintended state changes, financial loss, and deviation from expected contract logic. The affected component is the contract state, particularly storage variables and ledger entries influenced by user-provided parameters. The failure is \textit{irreversible} because once invalid inputs are accepted and committed on-chain, their effects become permanent and cannot really be automatically corrected.
Commonly referred to as ``Improper Validation''~\cite{kushwaha2022systematic}.

\subsubsection{\textbf{Undetected Sub-call Failure}}
This failure occurs when a smart contract executes an external sub-call that fails, but the return value or exception is not checked, letting execution proceed as if the operation succeeded~\cite{kushwaha2022systematic}. It can be exploited to propagate incorrect logic or to trigger denial-of-service attacks through repeated failed calls in loops~\cite{wu2024comprehensive}. It originates in the smart contract layer and is caused by improper handling of external call results (missing return-value checks, inadequate exception propagation); the root cause lies in flawed integration with external contracts, where failed sub-calls are not detected or handled, allowing execution to continue under incorrect assumptions.
\textit{Integrity} is violated, as state transitions rely on incorrect assumptions about external operation success. The impacts include incorrect downstream state updates, incomplete or inconsistent execution, silent error propagation, and unreliable system behavior. The affected component is the contract execution logic, particularly paths dependent on external call results and inter-contract interactions. The failure is \textit{transient}, as it affects specific execution paths and can be mitigated through proper validation and error handling in future executions.
Commonly referred to as ``Improper Exception Handling''~\cite{kushwaha2022systematic}.

\subsubsection{\textbf{Identifier Uniqueness Violation}}
This failure occurs when distinct entities are treated as identical due to non-unique or colliding identifiers, breaking referential integrity~\cite{kushwaha2022systematic}. A common example is the misuse of \texttt{abi.encodePacked()} in Solidity, where different inputs can produce identical byte sequences, enabling identifier collisions and exploitation of the contract's identification logic. It originates in the smart contract layer and is caused by identifier management and encoding design flaws where uniqueness constraints are not enforced during data representation or hashing; the root cause lies in unsafe encoding practices that allow different inputs to generate identical identifiers.
\textit{Authenticity} is violated, as the system can no longer reliably distinguish or verify unique entities. The impacts include identity confusion, broken referential integrity, incorrect entity associations, data corruption, collision-based exploitation, and loss of trust. The affected component is the contract state, particularly mappings, registries, and reference structures used for entity tracking. The failure is \textit{irreversible} because once identifiers collide, and the state is updated, the resulting corruption of relationships cannot really be automatically recovered.
Commonly referred to as ``Hash Collision''~\cite{kushwaha2022systematic}.

\subsubsection{\textbf{Insufficient Consensus Enforcement}}
This failure occurs when critical operations bypass the required multi-signature quorum, allowing actions to execute without sufficient collective authorization~\cite{kushwaha2022systematic}. Beyond unauthorized execution, it can induce liveness violations where essential functions become blocked by unmet consensus conditions~\cite{stephens2021smartpulse}. It originates in the smart contract, where the multi-party approval logic, such as multi-signature or governance quorum, is incorrectly implemented or insufficiently enforced; the collective authorization here is a contract-level mechanism rather than the blockchain consensus protocol. The root cause lies in flawed approval logic (missing quorum checks, improper aggregation of approvals, incorrect enforcement of multi-signature policies).
\textit{Integrity} is violated, as operations can execute without satisfying required collective authorization constraints. The impacts include unauthorized execution of privileged actions, violation of governance rules, financial loss, and loss of trust. The affected component is the contract execution and governance layer, particularly logic governing multi-signature verification and asset-controlled operations. The failure is \textit{irreversible} because once insufficiently authorized actions are committed on-chain, their effects are permanently recorded and cannot be undone without external intervention.
Commonly referred to as ``Improper Access Control''~\cite{kushwaha2022systematic}.

Four of the five security properties admitted by the taxonomy are instantiated by the failures above (Confidentiality, Integrity, Availability, Authenticity). Non-Repudiation is available as a dimension value but is not exhibited as a primary violation in the literature set; whether this reflects a boundary of smart contract failure behavior or a scope effect of our literature set is an open question.

\azul{A failure is labeled by its dominant observable effect, the deviation a user or a dependent contract actually sees, rather than by its root cause, so a single event maps to one mode even when several mechanisms contribute. The modes are therefore not a mutually exclusive partition of causes, and a few pairs are naturally adjacent, for example \textit{Corrupted Value Computation} and \textit{Distorted Economic Output} (a wrong number written to state versus a wrong price derived from it), \textit{Duplicated Fund Disbursement} and \textit{Unauthorized Action Execution} (an extra payout from a single authorization versus an action by an illegitimate actor), and \textit{Incorrect External Data Ingestion} and \textit{Distorted Economic Output} (bad oracle data committed to state versus a manipulated price feeding a trade). In each such case, the mode is the one that best matches the effect that dominates the observed outcome.}

\subsection{Saturation Exercise}
\label{SaturationExercise}

Following Kundisch et al.'s guidance on post-construction evaluation~\cite{kundisch2022update}, we subject the taxonomy to a saturation exercise with three parts. \azul{The first applies the taxonomy to a systematic literature review carried out after construction, with an explicit search query and screening by venue and topic, and reported following the PRISMA guidelines~\cite{page2021prisma}.} The second maps each entry of the OpenSCV vulnerability taxonomy~\cite{vidal2024openscv} to the failure mode that captures its observable effect, checking that existing vulnerability catalogs are covered. The third classifies failure events that occurred after the literature search was closed, checking the taxonomy's stability on previously unseen failures.

\azul{The \textit{systematic literature review} queried Google Scholar with \texttt{(("smart contract" OR "smart contracts" OR "blockchain") AND ("failure mode" OR "failure model" OR "failure taxonomy" OR "failure classification")) OR ("blockchain failure" OR "blockchain failures" OR "smart contract failure" OR "smart contract failures")}, retrieving 6{,}279 records.}

\begin{figure}[t]
\centering
\includegraphics[width=\columnwidth]{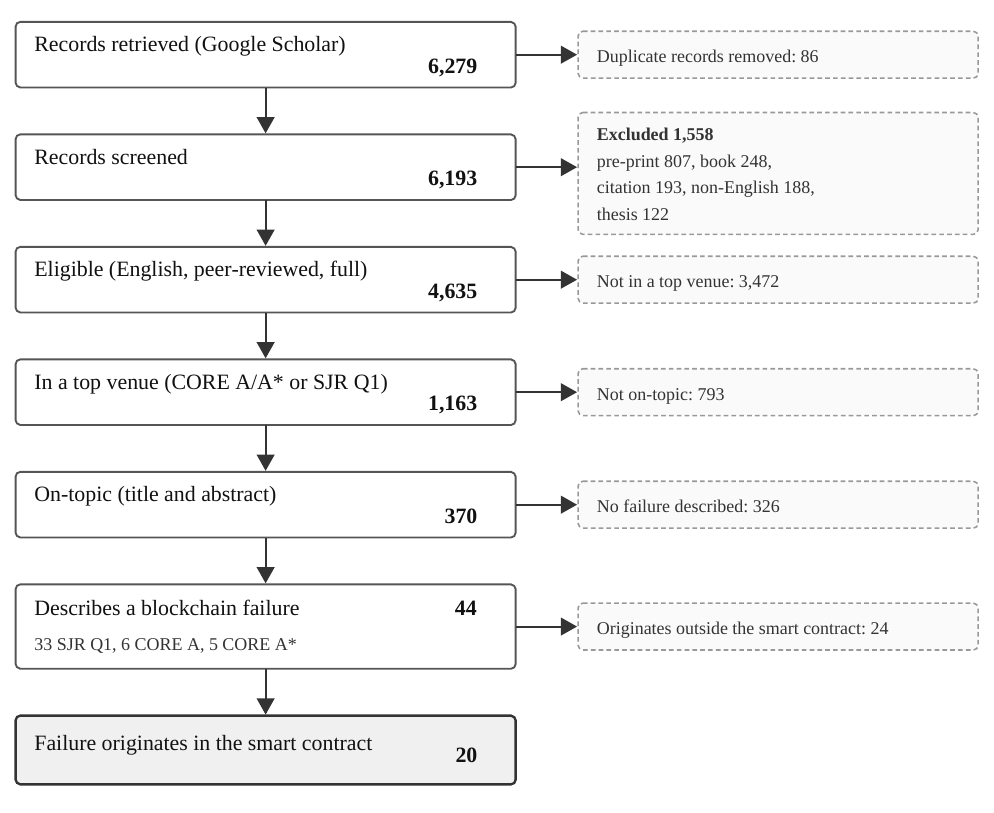}
\caption{\azul{PRISMA-style flow of the systematic literature review~\cite{page2021prisma}.}}
\label{fig:slrflow}
\end{figure}

\azul{Figure~\ref{fig:slrflow} shows the study selection flow. After removing 86 duplicate records, 6{,}193 were screened; screening removed 1{,}558 that were not English peer-reviewed articles with a full record, leaving 4{,}635, of which 1{,}163 appeared in a CORE A/A* or SJR Q1 venue. Of these, 370 were classified as on-topic based on their titles and abstracts, and full-text reading identified 44 that describe blockchain failures. An early-stage researcher and an experienced researcher then examined these 44 and, after resolving disagreements through discussion, retained the 20 whose failures originate in the smart contract; the remaining 24 describe failures rooted in other layers, such as consensus, sharding, or off-chain learning. Each of the 20 failures maps to an existing mode, and none required a new mode, dimension, or characteristic. The per-paper mapping and the full screening record are in the supplementary material~\cite{zenodo}.}

The \textit{OpenSCV coverage} check maps each entry of OpenSCV~\cite{vidal2024openscv}, \azul{a vulnerability taxonomy from our earlier work,} to the failure mode of the present taxonomy that captures its observable effect. The result of this mapping is reported in the \textit{OpenSCV Classification} column of the failure taxonomy, and the full entry-by-entry mapping is available in the supplementary material~\cite{zenodo}. Every OpenSCV entry was successfully mapped to one of the failure modes in our taxonomy; no entry produced an observable failure mode that required introducing a new category or dimension. \azul{The entries distribute across 21 of the 23 modes, concentrating in \textit{Inconsistent State Transition}, \textit{Unauthorized Action Execution}, \textit{Wasted Execution}, and \textit{Corrupted Value Computation}; only \textit{Network Throughput Degradation} and \textit{Insufficient Consensus Enforcement} receive no entry, consistent with their consensus- and network-level character.}

The check on \textit{external failure events} starts from the cutoff of the literature search. The most recent paper used to identify a failure mode is \cite{wang2025one}, which was published as an arXiv preprint in November 2025. We therefore take December 2025 as a cutoff and consider smart contract failure events reported after that date in two sources: \textit{Rekt News} (\url{https://rekt.news/}), a curated repository of DeFi incident post-mortems, and \textit{DeFiLlama Hacks} (\url{https://defillama.com/hacks}), a structured incident database.
\vermelho{We reviewed every Ethereum event reported by these sources from December 2025 onward and until the moment of writing, using DeFiLlama Hacks as the enumeration source because it is known to be quite complete (in terms of the number of registered incidents), and Rekt News, together with security-firm post-mortems, to determine how each incident occurred.}

\azul{Cross-checking the two sources also corrected one label: an event that DeFiLlama records as a private-key compromise is, according to its Rekt post-mortem, a smart-contract access-control failure, and was reinstated in scope. Of the 61 Ethereum events after the cutoff, 12 were out of scope because their origin was not the smart contract (private-key, admin-key, hot-wallet, DNS, social-engineering, or supply-chain compromise).}

\azul{Each of the remaining in-scope events was classified by its observable effect, the majority of them by two reviewers working in joint discussion; an event was not assigned a mode when no reliable post-mortem existed to establish the mechanism. Six events were excluded for insufficient public information on the mechanism, three of them reported only through a tracker technique label with no post-mortem to confirm the mechanism; and one, a Sybil reward-farming abuse in which the reward contract executed as designed, was excluded as out of scope. The resulting 42 classified events form the saturation set; the full annotated list, with a source for each event and the two out-of-scope groups, is in the supplementary material~\cite{zenodo}.}

\begin{table}[htbp]
\centering
\caption{\azul{Failure modes exercised by the 42 in-scope post-cutoff events.}}
\label{tab:saturationcoverage}
\renewcommand{\arraystretch}{1.0}
\setlength{\tabcolsep}{6pt}
\footnotesize
\begin{tabular}{@{}l r@{}}
\hline
\textbf{Failure mode} & \textbf{Events} \\
\hline
Unauthorized Action Execution & 18 \\
Distorted Economic Output & 11 \\
Duplicated Fund Disbursement & 5 \\
Insufficient Consensus Enforcement & 2 \\
Corrupted Value Computation & 2 \\
Incorrect External Data Ingestion & 2 \\
Misdirected Token Transfer & 1 \\
Inconsistent State Transition & 1 \\
\hline
\textbf{Total} & \textbf{42} \\
\hline
\end{tabular}
\end{table}

\azul{The 42 events mapped to eight failure modes already present in the taxonomy (Table~\ref{tab:saturationcoverage}), led by Unauthorized Action Execution (18) and Distorted Economic Output (11). No event required a new failure mode, dimension, or characteristic, and every dimension value used was already present among the 23 modes. With no materiality threshold, the evidence spans a large and varied set of incidents rather than only a handful of high-value ones, and the taxonomy remained stable across it. The events still concentrate in a subset of the modes, because public trackers report financially material exploits rather than the availability, privacy, or execution-integrity failures that the OpenSCV mapping does exercise; this reflects reporting bias in the sources rather than a gap in the taxonomy. Every OpenSCV entry likewise maps to a failure mode, confirming that existing vulnerability catalogs are covered.}

\section{CIPER: A Failure Model for Smart Contracts}
\label{sec:ciper}

\azul{This section derives the CIPER failure model from the taxonomy and then applies it to five documented failures, one for each severity class, showing how a failure is read from its observable effect to a failure mode and, from there, to a severity class.}

\subsection{The CIPER Model}

We condense the failures identified in our taxonomy into a failure model by grouping them according to observable behavior and severity. To do this, we analyzed each failure along its impact-related dimensions: affected components, violated security properties, impact propagation, and whether effects are transient or irreversible. The resulting classification is inspired by the CRASH failure model~\cite{koopman2000ballista}, which categorizes failures by observable impact rather than origin, but is adapted to the specifics of blockchain systems, where immutability and deterministic state replication mean that once a faulty state transition is committed, it is globally replicated and effectively irreversible. We therefore treat irreversibility, state persistence, and direct financial impact as primary severity drivers; failures that only disrupt execution or degrade performance are comparatively less severe, since they remain localized, transient, and recoverable through re-execution or parameter adjustment. The resulting scale spans, from most to least severe, irreversible loss of state or funds, persistent incorrectness or privacy violation, transaction-level execution disruption, performance degradation, and service unavailability.

Table~\ref{tab:ciper} presents the \textbf{CIPER} failure model. Each row describes a severity class through its name and definition, the nature of failures within the class, typical impacts, the main components affected, the violated security property, and the failure modes from the taxonomy that fall under the class. The mapping of every failure mode to its CIPER class is available in the supplementary material~\cite{zenodo}.

A failure of the taxonomy is assigned to a severity class according to the following rules, applied in order:
\begin{enumerate}[label=\arabic*)]
    \item \textbf{Critical (C)} if the failure directly corrupts the ledger (balances, token ownership) or causes irreversible loss of funds through unauthorized transfers.
    \item \textbf{High (I)} if the failure compromises the integrity, confidentiality, or authenticity of non-ledger state (pricing, privacy, identity, authorization) without directly corrupting the ledger.
    \item \textbf{Medium (P)} if the failure disrupts execution at the transaction level, producing incomplete or inconsistent operations with typically transient effects.
    \item \textbf{Low (E)} if the failure degrades performance or resource efficiency without corrupting state or blocking access.
    \item \textbf{Minimal (R)} if the failure makes contract functionality inaccessible while the underlying state remains intact.
\end{enumerate}

\azul{These rules rank failures by their technical impact on ledger state, namely the extent of on-chain state corruption, its propagation potential, and its irreversibility, rather than by the magnitude of financial loss or the preference of any particular stakeholder.}

\begin{table*}[!t]
\centering
\caption{The CIPER failure model: severity levels and their characteristics.}
\label{tab:ciper}
\renewcommand{\arraystretch}{0.9}
\setlength{\tabcolsep}{2pt}
\scriptsize
\begin{tabular}{@{}>{\raggedright\arraybackslash}p{0.9cm} >{\raggedright\arraybackslash}p{1.7cm} >{\raggedright\arraybackslash}p{1.8cm} >{\raggedright\arraybackslash}p{2.7cm} >{\raggedright\arraybackslash}p{1.9cm} >{\raggedright\arraybackslash}p{1.6cm} >{\raggedright\arraybackslash}p{1.5cm} >{\raggedright\arraybackslash}p{4.6cm}@{}}
\hline
\textbf{Severity} & \textbf{Failure Class} & \textbf{Class Definition} & \textbf{Description} & \textbf{Impacts} & \textbf{Main Components Affected} & \textbf{Security Property} & \textbf{Failure Modes} \\
\hline
Critical & {\footnotesize\textbf{\textcolor{darkred}{C}}}ritical State Compromise & Irreversible Financial \& State Compromise & Failures causing permanent loss of assets or irreversible corruption of blockchain state & Fund loss, duplicated spending, invalid ledger, unauthorised state changes & State (ledger, balances, assets) & Integrity; Authenticity & \textit{Double-Spending}; \textit{Incorrect External Data Ingestion}; \textit{Corrupted Value Computation}; \textit{Inconsistent State Transition}; \textit{Misdirected Token Transfer}; \textit{Duplicated Fund Disbursement}; \textit{Unauthorized Action Execution}; \textit{Insufficient Consensus Enforcement} \\
\hline
High & {\footnotesize\textbf{\textcolor{darkred}{I}}}ntegrity \& Privacy Violation & Persistent Incorrectness, Privacy, or Authorisation Violation & Failures causing lasting or recurrent incorrectness, privacy exposure, or unauthorised identity/access without immediate total loss & Incorrect state, data leakage, privacy breach, corrupted computations, unauthorised access & State, Accounts, Identity & Integrity; Confidentiality; Authenticity & \textit{Transaction History Disclosure}; \textit{Unintended State Visibility}; \textit{Identifier Uniqueness Violation}; \textit{Distorted Economic Output}; \textit{Stale State Operation} \\
\hline
Medium & {\footnotesize\textbf{\textcolor{darkred}{P}}}rocess Execution Failure & Execution Integrity Degradation & Failures affecting correctness and completeness of execution; effects are often transient but may persist when invalid inputs are committed & Failed workflows, inconsistent execution, silent errors & Execution; state transitions & Integrity & \textit{Premature Execution Halt}; \textit{Silent Operation Abort}; \textit{Nested Execution Collapse}; \textit{Undetected Sub-call Failure}; \textit{Unchecked Input Acceptance} \\
\hline
Low & {\footnotesize\textbf{\textcolor{darkred}{E}}}fficiency \& Performance Degradation & Resource \& Performance Inefficiency or Degradation & Failures that waste resources or degrade performance without corrupting state & Gas waste, latency, throughput reduction & Network; computation resources & Availability & \textit{Wasted Execution}; \textit{Network Throughput Degradation} \\
\hline
Minimal & {\footnotesize\textbf{\textcolor{darkred}{R}}}eliability \& Availability Disruption & Service Reliability or Availability Disruption & Failures preventing access to system functionality without necessarily corrupting state & Blocked functions, unusable contract, service downtime & Functionality; service layer & Availability & \textit{Persistent Unavailability}; \textit{Involuntary Contract Termination}; \textit{Permanent Asset Lock} \\
\hline
\end{tabular}
\end{table*}
Severity decreases from Critical (C) to Minimal (R). Although availability disruption may feel severe from a user perspective, R ranks below the state-corrupting classes because no on-chain value is lost, and service can be restored through redeployment or governance action without state recovery. \azul{Under a financial-loss lens, some placements are counterintuitive. \textit{Distorted Economic Output} is High rather than Critical because it distorts values derived from execution without directly corrupting the ledger, and \textit{Permanent Asset Lock} is Minimal because the ledger remains correct and only accessibility is lost. A ranking by financial or user impact would place these modes higher.}

\subsection{Applying the Model}

\azul{To illustrate the assignment, we take one documented failure for each CIPER class and classify it as we did the saturation events, deriving the dominant failure mode from the observable effect and then applying the ordered rules. The Critical and High cases are drawn from the post-cutoff incident set; the Medium, Low, and Minimal cases, whose classes public trackers rarely report because they carry little or no direct financial loss, are drawn from the smart contract literature. Table~\ref{tab:ciperexamples} summarizes the five, and the readings follow.}

\begin{table}[!t]
\centering
\caption{\azul{Five documented failures, one per CIPER class, read from observable effect to failure mode and severity class.}}
\label{tab:ciperexamples}
\renewcommand{\arraystretch}{1.15}
\setlength{\tabcolsep}{4pt}
\footnotesize
\begin{tabular}{@{}>{\raggedright\arraybackslash}p{2.0cm} >{\raggedright\arraybackslash}p{2.9cm} >{\raggedright\arraybackslash}p{1.4cm} >{\raggedright\arraybackslash}p{0.7cm}@{}}
\hline
\textbf{Case} & \textbf{Failure mode} & \textbf{CIPER class} & \textbf{Src.} \\
\hline
Truebit & \textit{Corrupted Value Computation} & (C) Critical & \cite{rekt_truebit} \\
YO Protocol & \textit{Distorted Economic Output} & (I) High & \cite{rekt_yoprotocol} \\
King of the Ether Throne & \textit{Undetected Sub-call Failure} & (P) Medium & \cite{atzei2017survey} \\
Ineffective (dead-code) execution & \textit{Wasted Execution} & (E) Low & \cite{hajdu2020using} \\
Parity wallet freeze & \textit{Permanent Asset Lock} & (R) Minimal & \cite{atzei2017survey} \\
\hline
\end{tabular}
\end{table}

\azul{In the Truebit incident of January 2026~\cite{rekt_truebit}, an unchecked addition in the bonding-curve pricing overflowed, so the purchase-price function returned zero and the contract minted tokens for almost nothing. A numeric computation wrote an incorrect value that propagated through contract state, which is \textit{Corrupted Value Computation}; \textit{Integrity} is violated and the effect is irreversible, so Rule~1 places it in \textit{Critical State Compromise}. In YO Protocol~\cite{rekt_yoprotocol}, an automated harvest swap ran with its slippage guard effectively disabled and routed the vault's position through thin pools, returning almost nothing; the transfers were authorized, and the ledger recorded them correctly, so Rule~1 does not apply. The unvalidated input quote is the cause, but the taxonomy assigns a mode by the dominant observable effect, a trade settled at a grossly distorted price, which is \textit{Distorted Economic Output}, placed by Rule~2 in \textit{Integrity \& Privacy Violation}.}

\azul{In King of the Ether Throne~\cite{atzei2017survey}, the payment to the previous king used \texttt{send}, and the failed sub-call was not checked, so execution continued as if the payment had completed; Rules~1 and~2 do not apply, and this transaction-level disruption is an \textit{Undetected Sub-call Failure}, which Rule~3 places in \textit{Process Execution Failure}. A contract that runs redundant or dead code on every call~\cite{hajdu2020using} consumes gas without changing any state; the transaction still completes and commits its intended effect, so nothing is corrupted, halted, or blocked and only resources are wasted, which is \textit{Wasted Execution} and, by Rule~4, \textit{Efficiency \& Performance Degradation}. Finally, the Parity multisig freeze~\cite{atzei2017survey} left the wallets' funds permanently inaccessible after a library self-destruct, while the ledger remained correct; earlier rules do not apply, so this \textit{Permanent Asset Lock} falls under Rule~5, \textit{Reliability \& Availability Disruption}.}

\azul{The five span the full scale and show that CIPER orders by observable impact on ledger state rather than by the amount at stake; the Parity freeze, for instance, immobilized far more value than the Truebit incident, yet ranks Minimal because no on-chain value was corrupted and only access was lost.}

\section{Threats to Validity}
\label{sec:threats}

Several factors may affect the validity of the results. First, the taxonomy is derived from a literature review prioritizing peer-reviewed research; failure modes reported only in practitioner forums or closed-source post-mortems may not be represented. \azul{The saturation exercise mitigates this in part. Its systematic literature review applies a search broader than the one used in construction, and its mapping of the OpenSCV vulnerability catalog covers the observable outcomes of a broad body of vulnerability literature; neither yielded a failure mode absent from the taxonomy. The construction and saturation searches also used failure-oriented terms and may still miss papers that describe the same failures under vulnerability, exploit, attack, or bug terminology.}

Second, the assignment of failure modes to dimensions and the grouping of failure modes into CIPER classes involved judgment by the three authors; although disagreements were resolved through discussion at each iteration, independent annotators from outside the authorship could produce different mappings for borderline cases. \azul{We did not compute independent inter-rater agreement; instead, the per-iteration outputs are published in full~\cite{zenodo}, giving a traceable record of how each assignment evolved. The taxonomy assigns each event its dominant observable failure mode rather than a mutually exclusive label, so adjacent modes may share causes and some borderline events admit alternative classifications.}

Third, the saturation exercise complements the literature review with post-cutoff incidents drawn from public incident trackers; \azul{although we applied no materiality threshold and considered events of any size, the set is still limited to publicly reported incidents, which concentrate in financially material exploits, so availability, privacy, and other less-reported failure modes may remain underrepresented in this evidence.}

Fourth, the scope is explicitly Ethereum and EVM-compatible smart contracts; chains using different programming models (for example, Cadence on Flow or Move-based chains) may exhibit failure modes that the taxonomy does not yet capture.

Finally, the \textit{Security Property} dimension admits Non-Repudiation as an additional value, but no failure in the literature set instantiates it, and the \textit{Security Principles Coverage} ending condition was intentionally defined to require only the three properties of the CIA triad rather than the full set. Whether the absence of Non-Repudiation reflects a boundary of smart contract failure behavior or a scope effect of the literature set is itself a threat to the completeness of the \textit{Security Property} dimension.

\section{Conclusion}
\label{sec:conclusion}

This paper presented a failure mode taxonomy for Ethereum smart contracts, developed from a literature review through five iterations of the Nickerson method. The taxonomy organizes observable failures of deployed smart contracts along five dimensions, covering their root cause, violated security property, impact (primary, secondary, and other), affected component, and the nature of the failure. From the taxonomy, we derived the CIPER failure model, which classifies failures into five severity levels, from irreversible state or financial compromise down to service unavailability. \azul{A saturation exercise with three parts: a systematic literature review carried out after construction, a mapping of the effects of every OpenSCV vulnerability entry to the failure modes, and the classification of 42 post-cutoff incidents drawn from public incident trackers, produced no new failure modes or dimensions.} The taxonomy, therefore, offers a shared vocabulary for describing what smart contracts do wrong when they fail, and the CIPER model links that vocabulary to the severity of each observable outcome.

\bibliographystyle{IEEEtran}
\bibliography{references/library,references/nuno}

@article{mukherjee2025doublespending,
  title = {Double-Spending Attacks in Cross-Blockchain Ecosystems},
  author = {Mukherjee, Aradhita and Olivieri, Luca and Chaki, Nabendu and Cortesi, Agostino},
  journal = {Blockchain: Research and Applications},
  pages = {100378},
  year = {2025},
  month = sep,
  issn = {2096-7209},
  doi = {10.1016/j.bcra.2025.100378},
  url       = {https://doi.org/10.1016/j.bcra.2025.100378},
  publisher = {Elsevier}
}

@inproceedings{frowis2023detecting,
  title={Detecting Privileged Parties on Ethereum},
  author={Fr{\"o}wis, Michael and B{\"o}hme, Rainer},
  booktitle={Financial Cryptography and Data Security. FC 2023 International Workshops},
  pages={470--488},
  year={2024},
  publisher={Springer},
  doi={10.1007/978-3-031-48806-1_30},
  url={https://doi.org/10.1007/978-3-031-48806-1_30}
}

@article{chen2020survey,
  title={A Survey on Ethereum Systems Security: Vulnerabilities, Attacks, and Defenses},
  author={Chen, Huashan and Pendleton, Marcus and Njilla, Laurent and Xu, Shouhuai},
  journal={ACM Computing Surveys (CSUR)},
  volume={53},
  number={3},
  articleno={67},
  numpages={43},
  year={2020},
  publisher={ACM},
  doi={10.1145/3391195},
  url       = {https://doi.org/10.1145/3391195},
}

@inproceedings{chen2019tokenscope,
  author    = {Chen, Ting and Zhang, Yufei and Li, Zihao and Luo, Xiapu and Wang, Ting and Cao, Rong and Xiao, Xiuzhuo and Zhang, Xiaosong},
  title     = {TokenScope: Automatically Detecting Inconsistent Behaviors of Cryptocurrency Tokens in Ethereum},
  booktitle = {ACM SIGSAC Conference on Computer and Communications Security (CCS)},
  year      = {2019},
  pages     = {1503--1520},
  doi       = {10.1145/3319535.3345664}
}

@article{ma2023pied,
  title={Pied-Piper: Revealing the Backdoor Threats in Ethereum ERC Token Contracts},
  author={Ma, Fuchen and Ren, Meng and Ouyang, Lerong and Chen, Yuanliang and Zhu, Juan and Chen, Ting and Zheng, Yingli and Dai, Xiao and Jiang, Yu and Sun, Jiaguang},
  journal={ACM Transactions on Software Engineering and Methodology},
  volume={32},
  number={3},
  articleno={61},
  numpages={24},
  year={2023},
  publisher={ACM},
  doi={10.1145/3560264},
  url       = {https://doi.org/10.1145/3560264},
}

@inproceedings{khan2023dynamic,
  title={Dynamic Analysis for Detection of Self-Destructive Smart Contracts},
  author={Khan, Zulfiqar Ali and Namin, Akbar Siami},
  booktitle={2023 IEEE 47th Annual Computers, Software, and Applications Conference (COMPSAC)},
  pages={1093--1100},
  year={2023},
  organization={IEEE},
  doi={10.1109/COMPSAC57700.2023.00165},
  url       = {https://doi.org/10.1109/COMPSAC57700.2023.00165},
}

@inproceedings{sun2024doubleup,
  title     = {DoubleUp Roll: Double-spending in Arbitrum by Rolling It Back},
  author    = {Sun, Zhiyuan and Li, Zihao and Peng, Xinghao and Luo, Xiapu and Jiang, Muhui and Zhou, Hao and Zhang, Yang},
  booktitle = {ACM SIGSAC Conference on Computer and Communications Security},
  series    = {CCS '24},
  year      = {2024},
  pages     = {2577--2590},
  publisher = {ACM},
  doi       = {10.1145/3658644.3690256},
  url       = {https://doi.org/10.1145/3658644.3690256},
}

@misc{eip6780,
  author       = {Guillaume Ballet and Vitalik Buterin and Dankrad Feist},
  title        = {{EIP-6780}: {SELFDESTRUCT} only in same transaction},
  year         = {2023},
  howpublished = {\url{https://eips.ethereum.org/EIPS/eip-6780}},
  note         = {Ethereum Improvement Proposals}
}

@inproceedings{nikolic2018finding,
  author    = {Ivica Nikoli{\'{c}} and Aashish Kolluri and Ilya Sergey and Prateek Saxena and Aquinas Hobor},
  title     = {Finding The Greedy, Prodigal, and Suicidal Contracts at Scale},
  booktitle = {34th Annual Computer Security Applications Conference ({ACSAC} 2018)},
  year      = {2018},
  pages     = {653--663},
  publisher = {{ACM}},
  doi       = {10.1145/3274694.3274743},
  url       = {https://doi.org/10.1145/3274694.3274743},
}

@article{hajdu2020using,
  author={Hajdu, {\'A}kos and Ivaki, Naghmeh and Kocsis, Imre and Klenik, Attila and G{\"o}nczy, L{\'a}szl{\'o} and Laranjeiro, Nuno and Madeira, Henrique and Pataricza, Andr{\'a}s},
  journal={IEEE Access},
  title={Using Fault Injection to Assess Blockchain Systems in Presence of Faulty Smart Contracts},
  year={2020},
  doi={10.1109/ACCESS.2020.3032239},
  url       = {https://doi.org/10.1109/ACCESS.2020.3032239},
}

@article{buterin2024privacy,
  author    = {Vitalik Buterin and Jacob Illum and Matthias Nadler and Fabian Sch\"{a}r and Ameen Soleimani},
  title     = {Blockchain privacy and regulatory compliance: Towards a practical equilibrium},
  journal   = {Blockchain: Research and Applications},
  volume    = {5},
  number    = {1},
  pages     = {100176},
  year      = {2024},
  issn      = {2096-7209},
  publisher = {Elsevier},
  doi       = {10.1016/j.bcra.2023.100176},
  url       = {https://doi.org/10.1016/j.bcra.2023.100176}
}

@inproceedings{bose2022sailfish,
  title={SAILFISH: vetting smart contract state-inconsistency bugs in seconds},
  author={Bose, P. and Das, D. and Chen, Y. and others},
  booktitle={2022 IEEE Symposium on Security and Privacy (SP)},
  pages={161--178},
  year={2022},
  organization={IEEE},
  doi={10.1109/SP46214.2022.9833721},
  url       = {https://doi.org/10.1109/SP46214.2022.9833721},
}

@article{vidal2025analyzing,
  title={Analyzing the impact of elusive faults on blockchain reliability},
  author={Vidal, Fernando Richter and Ivaki, Naghmeh and Laranjeiro, Nuno},
  journal={Blockchain: Research and Applications},
  volume={6},
  pages={100295},
  year={2025},
  publisher={Elsevier},
  doi={10.1016/j.bcra.2025.100295},
  url       = {https://doi.org/10.1016/j.bcra.2025.100295},
}

@article{ma2023vgas,
  author = {Ma, Fuchen and Ren, Meng and Ying, Fu and Sun, Wanting and Song, Houbing and Shi, Heyuan and Jiang, Yu and Li, Huizhong},
  title = {V-Gas: Generating High Gas Consumption Inputs to Avoid Out-of-Gas Vulnerability},
  journal = {ACM Transactions on Internet Technology},
  volume = {23},
  number = {3},
  articleno = {40},
  numpages = {22},
  year = {2023},
  month = {aug},
  publisher = {ACM},
  doi = {10.1145/3511900},
  url = {https://doi.org/10.1145/3511900}
}

@article{pacheco2023ethereum,
  title={What makes Ethereum blockchain transactions be processed fast or slow? An empirical study},
  author={Pacheco, Michael and Oliva, Gustavo A. and Rajbahadur, Gopi Krishnan and Hassan, Ahmed E.},
  journal={Empirical Software Engineering},
  volume={28},
  number={2},
  pages={39},
  year={2023},
  publisher={Springer},
  doi={10.1007/s10664-022-10283-7},
  url={https://doi.org/10.1007/s10664-022-10283-7}
}

@inproceedings{munir2023statically,
  title={Statically Checking Missing Input Validations in Solidity Smart Contracts - A Case Study},
  author={Munir, Sundas and Baig, Mirza Sanam Iqbal and Noor, Mah and Murad, Syeda Hina},
  booktitle={2023 IEEE International Conference on Blockchain (Blockchain)},
  pages={47--54},
  year={2023},
  organization={IEEE},
  doi={10.1109/Blockchain60715.2023.00017},
  url={https://doi.org/10.1109/Blockchain60715.2023.00017}
}

@misc{buterin2014ethereum,
  author       = {Buterin, Vitalik},
  title        = {{Ethereum}: A Next-Generation Smart Contract and Decentralized Application Platform},
  year         = {2014},
  howpublished = {White paper},
  url          = {https://ethereum.org/content/whitepaper/whitepaper-pdf/Ethereum_Whitepaper_-_Buterin_2014.pdf}
}

@inproceedings{vidal2025bbench,
  author    = {Vidal, Fernando Richter and Ivaki, Naghmeh and Laranjeiro, Nuno},
  title     = {{bBench}: A Comprehensive Performance Benchmark for Blockchain Applications},
  booktitle = {36th IEEE International Symposium on Software Reliability Engineering (ISSRE)},
  year      = {2025},
  month     = {October},
  address   = {S\~{a}o Paulo, Brazil},
  publisher = {IEEE},
  pages     = {382--393},
  doi       = {10.1109/ISSRE66568.2025.00046},
  url       = {https://doi.org/10.1109/ISSRE66568.2025.00046}
}

@inproceedings{torres2018osiris,
  author    = {Torres, Christof Ferreira and Sch\"{u}tte, Julian and Mathis, Radu},
  title     = {OSIRIS: Hunting for Integer Bugs in Ethereum Smart Contracts},
  booktitle = {34th Annual Computer Security Applications Conference (ACSAC)},
  year      = {2018},
  pages     = {664--676},
  publisher = {ACM},
  doi       = {10.1145/3274694.3274737},
  url       = {https://doi.org/10.1145/3274694.3274737},
}

@article{kushwaha2022systematic,
  title={Systematic Review of Security Vulnerabilities in Ethereum Blockchain Smart Contract},
  author={Kushwaha, Satpal Singh and Joshi, Sandeep and Singh, Dilbag and Kaur, Manjit and Lee, Heung-No},
  journal={IEEE Access},
  volume={10},
  pages={6605--6621},
  year={2022},
  publisher={IEEE},
  doi={10.1109/ACCESS.2021.3140091},
  url       = {https://doi.org/10.1109/ACCESS.2021.3140091},
}

@techreport{iso27000_2018,
  author = {{International Organization for Standardization}},
  type = {Standard},
  key = {ISO/IEC 27000:2018},
  year = {2018},
  title = {{ISO/IEC 27000:2018}: Information technology -- Security techniques -- Information security management systems -- Overview and vocabulary},
  institution = {International Organization for Standardization (ISO)},
  address = {Geneva, Switzerland},
  url = {https://www.iso.org/standard/73906.html}
}

@article{wu2024comprehensive,
  title={A comprehensive survey of smart contract security: State of the art and research directions},
  author={Wu, Guangfu and Wang, HaiPing and Lai, Xin and Wang, Mengmeng and He, Daojing and Chan, Sammy},
  journal={Journal of Network and Computer Applications},
  volume={226},
  pages={103882},
  year={2024},
  publisher={Elsevier},
  issn={1084-8045},
  doi={10.1016/j.jnca.2024.103882},
  url       = {https://doi.org/10.1016/j.jnca.2024.103882},
}

@inproceedings{stephens2021smartpulse,
  title={SmartPulse: Automated Checking of Temporal Properties in Smart Contracts},
  author={Stephens, Jon and Ferles, Kostas and Mariano, Benjamin and Lahiri, Shuvendu and Dillig, Isil},
  booktitle={2021 IEEE Symposium on Security and Privacy (SP)},
  pages={555--572},
  year={2021},
  organization={IEEE},
  doi={10.1109/SP40001.2021.00085}
}

@misc{wang2025one,
  author        = {Wang, Zexu and Chen, Jiachi and Lin, Zewei and Chen, Wenqing and Ning, Kaiwen and Yu, Jianxing and Feng, Yuming and Zhang, Yu and Zhang, Weizhe and Zheng, Zibin},
  title         = {One Signature, Multiple Payments: Demystifying and Detecting Signature Replay Vulnerabilities in Smart Contracts},
  year          = {2025},
  eprint        = {2511.09134},
  archivePrefix = {arXiv}
}

@inproceedings{perez2020broken,
  title={Broken Metre: Attacking Resource Metering in EVM},
  author={Perez, Daniel and Livshits, Benjamin},
  booktitle={27th Annual Network and Distributed System Security Symposium (NDSS)},
  year={2020},
  organization={Internet Society}
}

@inproceedings{li2023demystifying,
  title={Demystifying DeFi MEV Activities in Flashbots Bundle},
  author={Li, Zihao and Li, Jianfeng and He, Zheyuan and Luo, Xiapu and Wang, Ting and Ni, Xiaoze and Yang, Wenwu and Chen, Xi and Chen, Ting},
  booktitle={2023 ACM SIGSAC Conference on Computer and Communications Security},
  year={2023},
  publisher={ACM},
  address={New York, NY, USA},
  doi={10.1145/3576915.3616590}
}

@inproceedings{xi2024pomabuster,
  title={POMABuster: Detecting Price Oracle Manipulation Attacks in Decentralized Finance},
  author={Xi, Rui and Wang, Zehua and Pattabiraman, Karthik},
  booktitle={2024 IEEE Symposium on Security and Privacy (SP)},
  year={2024},
  organization={IEEE},
  doi={10.1109/SP54263.2024.00257}
}

@inproceedings{deng2024safeguarding,
  author = {Deng, Xun and Beillahi, Sidi Mohamed and Minwalla, Cyrus and Du, Han and Veneris, Andreas and Long, Fan},
  title = {Safeguarding DeFi Smart Contracts against Oracle Deviations},
  booktitle = {2024 IEEE/ACM 46th International Conference on Software Engineering (ICSE '24)},
  month = {April 14--20},
  year = {2024},
  address = {Lisbon, Portugal},
  publisher = {ACM},
  pages = {12},
  doi = {10.1145/3597503.3639225}
}

@article{hassan2023trust,
  title={From trust to truth: Advancements in mitigating the Blockchain Oracle problem},
  author={Hassan, Ammar and Makhdoom, Imran and Iqbal, Waseem and Ahmad, Awais and Raza, Asad},
  journal={Journal of Network and Computer Applications},
  volume={217},
  pages={103672},
  year={2023},
  publisher={Elsevier},
  issn={1084-8045},
  doi={10.1016/j.jnca.2023.103672}
}

@article{koopman2000ballista,
  author  = {Philip Koopman and John DeVale},
  title   = {The Exception Handling Effectiveness of {POSIX} Operating Systems},
  journal = {IEEE Transactions on Software Engineering},
  volume  = {26},
  number  = {9},
  pages   = {837-848},
  year    = {2000},
  doi     = {10.1109/32.877845},
  url       = {https://doi.org/10.1109/32.877845},
}

@inproceedings{gunawi2014cloudbugs,
  author    = {Haryadi S. Gunawi and Mingzhe Hao and Tanakorn Leesatapornwongsa and Tiratat Patana-anake and Thanh Do and Jeffry Adityatama and Kurnia J. Eliazar and Agung Laksono and Jeffrey F. Lukman and Vincentius Martin and Anang D. Satria},
  title     = {What Bugs Live in the Cloud? {A} Study of 3000+ Issues in Cloud Systems},
  booktitle = {ACM Symposium on Cloud Computing (SoCC)},
  pages     = {1-14},
  year      = {2014},
  doi       = {10.1145/2670979.2670986},
  url       = {https://doi.org/10.1145/2670979.2670986},
}

@inproceedings{humbatova2020dlfaults,
  author    = {Nargiz Humbatova and Gunel Jahangirova and Gabriele Bavota and Vincenzo Riccio and Andrea Stocco and Paolo Tonella},
  title     = {Taxonomy of Real Faults in Deep Learning Systems},
  booktitle = {ACM/IEEE 42nd International Conference on Software Engineering (ICSE)},
  pages     = {1110-1121},
  year      = {2020},
  doi       = {10.1145/3377811.3380395},
  url       = {https://doi.org/10.1145/3377811.3380395},
}

@article{chen2020defects,
  author  = {Jiachi Chen and Xin Xia and David Lo and John Grundy and Xiapu Luo and Ting Chen},
  title   = {Defining Smart Contract Defects on {Ethereum}},
  journal = {IEEE Transactions on Software Engineering},
  volume  = {48},
  number  = {1},
  pages   = {327-345},
  year    = {2022},
  doi     = {10.1109/TSE.2020.2989002},
  url       = {https://doi.org/10.1109/TSE.2020.2989002},
}

@article{soud2024fly,
  author  = {Majd Soud and Grischa Liebel and Mohammad Hamdaqa},
  title   = {A fly in the ointment: an empirical study on the characteristics of {Ethereum} smart contract code weaknesses},
  journal = {Empirical Software Engineering},
  volume  = {29},
  number  = {1},
  year    = {2024},
  doi     = {10.1007/s10664-023-10398-5},
  url       = {https://doi.org/10.1007/s10664-023-10398-5},
}

@misc{rezaei2025rootcause,
  author        = {Hadis Rezaei and Mojtaba Eshghie and Karl Andersson and Francesco Palmieri},
  title         = {{SoK}: Root Cause of \$1 Billion Loss in Smart Contract Real-World Attacks via a Systematic Literature Review of Vulnerabilities},
  year          = {2025},
  eprint        = {2507.20175},
  archivePrefix = {arXiv}
}

@inproceedings{zhou2023sokdefi,
  author    = {Liyi Zhou and Xihan Xiong and Jens Ernstberger and Stefanos Chaliasos and Zhipeng Wang and Ye Wang and Kaihua Qin and Roger Wattenhofer and Dawn Song and Arthur Gervais},
  title     = {{SoK}: Decentralized Finance {(DeFi)} Attacks},
  booktitle = {IEEE Symposium on Security and Privacy (SP)},
  pages     = {2444-2461},
  year      = {2023},
  doi       = {10.1109/SP46215.2023.10179435},
  url       = {https://doi.org/10.1109/SP46215.2023.10179435},
}

@article{laranjeiro2021robustness,
  author    = {Nuno Laranjeiro and Jo\~{a}o Agnelo and Jorge Bernardino},
  title     = {A Systematic Review on Software Robustness Assessment},
  journal   = {ACM Computing Surveys},
  volume    = {54},
  number    = {4},
  articleno = {89},
  numpages  = {65},
  year      = {2021},
  doi       = {10.1145/3448977},
  url       = {https://doi.org/10.1145/3448977},
}

@inproceedings{werner2022sok,
  author    = {Werner, Sam M. and Perez, Daniel and
               Gudgeon, Lewis and Klages-Mundt, Ariah and
               Harz, Dominik and Knottenbelt, William J.},
  title     = {{SoK}: Decentralized Finance ({DeFi})},
  booktitle = {4th ACM Conference on
               Advances in Financial Technologies (AFT)},
  pages     = {30--46},
  year      = {2022},
  publisher = {ACM},
  doi       = {10.1145/3558535.3559780},
  url       = {https://doi.org/10.1145/3558535.3559780},
}

@misc{wood2014ethereum,
  author    = {Wood, Gavin},
  title     = {Ethereum: A Secure Decentralised Generalised
               Transaction Ledger},
  year      = {2014},
  url       = {https://ethereum.github.io/yellowpaper/paper.pdf},
  note      = {Ethereum Yellow Paper}
}

@misc{defillama2026,
  author    = {{DefiLlama}},
  title     = {{DeFi} Dashboard: Total Value Locked},
  year      = {2026},
  url       = {https://defillama.com/}
}

@inproceedings{atzei2017survey,
  author    = {Atzei, Nicola and Bartoletti, Massimo and
               Cimoli, Tiziana},
  title     = {A Survey of Attacks on {Ethereum} Smart
               Contracts ({SoK})},
  booktitle = {6th International
               Conference on Principles of Security and
               Trust (POST)},
  series    = {Lecture Notes in Computer Science},
  volume    = {10204},
  pages     = {164--186},
  year      = {2017},
  publisher = {Springer},
  doi       = {10.1007/978-3-662-54455-6_8},
  url       = {https://doi.org/10.1007/978-3-662-54455-6_8},
}

@inproceedings{perez2021vulnerable,
  author    = {Perez, Daniel and Livshits, Benjamin},
  title     = {Smart Contract Vulnerabilities: Vulnerable
               Does Not Imply Exploited},
  booktitle = {30th USENIX Security
               Symposium (USENIX Security 21)},
  pages     = {1325--1341},
  year      = {2021},
  publisher = {USENIX Association},
  isbn      = {978-1-939133-24-3}
}

@article{mehar2019understanding,
  author  = {Mehar, Muhammad Izhar and Shier, Charles Louis and Giambattista, Alana and Gong, Elgar and Fletcher, Gabrielle and Sanayhie, Ryan and Kim, Henry M. and Laskowski, Marek},
  title   = {Understanding a Revolutionary and Flawed Grand Experiment in Blockchain: The {DAO} Attack},
  journal = {Journal of Cases on Information Technology},
  volume  = {21},
  number  = {1},
  pages   = {19--32},
  year    = {2019},
  doi     = {10.4018/JCIT.2019010102}
}

@inproceedings{destefanis2018smart,
  author    = {Destefanis, Giuseppe and Marchesi, Michele and Ortu, Marco and Tonelli, Roberto and Bracciali, Andrea and Hierons, Robert},
  title     = {Smart Contracts Vulnerabilities: A Call for Blockchain Software Engineering?},
  booktitle = {IEEE Int. Workshop on Blockchain Oriented Software Engineering (IWBOSE)},
  pages     = {19--25},
  year      = {2018},
  doi       = {10.1109/IWBOSE.2018.8327567}
}

@inproceedings{qin2021attacking,
  author    = {Qin, Kaihua and Zhou, Liyi and Livshits,
               Benjamin and Gervais, Arthur},
  title     = {Attacking the {DeFi} Ecosystem with Flash
               Loans for Fun and Profit},
  booktitle = {Financial Cryptography and Data Security
               (FC 2021)},
  series    = {Lecture Notes in Computer Science},
  volume    = {12674},
  pages     = {3--32},
  year      = {2021},
  publisher = {Springer},
  doi       = {10.1007/978-3-662-64322-8_1},
  url       = {https://doi.org/10.1007/978-3-662-64322-8_1},
}

@misc{owaspsctop10_2026,
  author    = {{OWASP Foundation}},
  title     = {{OWASP} Smart Contract Top~10: 2026},
  year      = {2026},
  url       = {https://scs.owasp.org/sctop10/}
}

@article{iuliano2026slr,
  author    = {Iuliano, Gerardo and Di Nucci, Dario},
  title     = {Smart Contract Vulnerabilities, Tools, and
               Benchmarks: An Updated Systematic Literature
               Review},
  journal   = {Journal of Systems and Software},
  year      = {2026},
  note      = {In press},
  doi       = {10.1016/j.jss.2026.112370},
  url       = {https://doi.org/10.1016/j.jss.2026.112370},
}

@article{vidal2024openscv,
  author    = {Vidal, Fernando Richter and Ivaki, Naghmeh
               and Laranjeiro, Nuno},
  title     = {{OpenSCV}: An Open Hierarchical Taxonomy for
               Smart Contract Vulnerabilities},
  journal   = {Empirical Software Engineering},
  volume    = {29},
  number    = {4},
  pages     = {101},
  year      = {2024},
  doi       = {10.1007/s10664-024-10446-8},
  url       = {https://doi.org/10.1007/s10664-024-10446-8},
}

@article{avizienis2004basic,
  author    = {Avi\v{z}ienis, Algirdas and Laprie,
               Jean-Claude and Randell, Brian and Landwehr,
               Carl},
  title     = {Basic Concepts and Taxonomy of Dependable and
               Secure Computing},
  journal   = {IEEE Transactions on Dependable and Secure
               Computing},
  volume    = {1},
  number    = {1},
  pages     = {11--33},
  year      = {2004},
  doi       = {10.1109/TDSC.2004.2},
  url       = {https://doi.org/10.1109/TDSC.2004.2},
}

@article{nickerson2013method,
  author    = {Nickerson, Robert C. and Varshney, Upkar and
               Muntermann, Jan},
  title     = {A Method for Taxonomy Development and Its
               Application in Information Systems},
  journal   = {European Journal of Information Systems},
  volume    = {22},
  number    = {3},
  pages     = {336--359},
  year      = {2013},
  doi       = {10.1057/ejis.2012.26},
  url       = {https://doi.org/10.1057/ejis.2012.26},
}

@article{chillarege1992odc,
  author    = {Chillarege, Ram and Bhandari, Inderpal S. and
               Chaar, Jarir K. and Halliday, Michael J. and
               Moebus, Diane S. and Ray, Bonnie K. and
               Wong, Man-Yuen},
  title     = {Orthogonal Defect Classification: A Concept
               for In-Process Measurements},
  journal   = {IEEE Transactions on Software Engineering},
  volume    = {18},
  number    = {11},
  pages     = {943--956},
  year      = {1992},
  doi       = {10.1109/32.177364},
  url       = {https://doi.org/10.1109/32.177364},
}

@inproceedings{ruggiero2024sok,
  author    = {Ruggiero, Claudia and Mazzini, Pietro and
               Coppa, Emilio and Lenti, Simone and Bonomi,
               Silvia},
  title     = {{SoK}: A Unified Data Model for Smart
               Contract Vulnerability Taxonomies},
  booktitle = {19th International
               Conference on Availability, Reliability and
               Security (ARES)},
  year      = {2024},
  publisher = {ACM},
  doi       = {10.1145/3664476.3664507},
  url       = {https://doi.org/10.1145/3664476.3664507},
}

@misc{zenodo,
  author       = {Vidal, Fernando Richter and Ivaki, Naghmeh and Laranjeiro, Nuno},
  title        = {A Failure Mode Taxonomy and Severity Model for Ethereum Smart Contracts - Supplementary Material},
  howpublished = {Zenodo},
  month        = apr,
  year         = 2026,
  doi          = {10.5281/zenodo.19728992},
  url          = {https://doi.org/10.5281/zenodo.19728992},
}

@article{vidal2024vulnerability,
  author  = {Vidal, Fernando Richter and Ivaki, Naghmeh and Laranjeiro, Nuno},
  title   = {Vulnerability detection techniques for smart contracts: A systematic literature review},
  journal = {Journal of Systems and Software},
  volume  = {217},
  pages   = {112160},
  year    = {2024},
  issn    = {0164-1212},
  doi     = {10.1016/j.jss.2024.112160},
  url     = {https://doi.org/10.1016/j.jss.2024.112160}
}

@article{kundisch2022update,
  author  = {Kundisch, Dennis and Muntermann, J{\"o}rg and Oberl{\"a}nder, Anna Maria and Rau, Daniel and R{\"o}glinger, Maximilian and Schoormann, Thorsten and Szopinski, Daniel},
  title   = {An Update for Taxonomy Designers: Methodological Guidance from Information Systems Research},
  journal = {Business \& Information Systems Engineering},
  volume  = {64},
  number  = {4},
  pages   = {421--439},
  year    = {2022},
  doi     = {10.1007/s12599-021-00723-x},
  url     = {https://doi.org/10.1007/s12599-021-00723-x}
}

@misc{rekt_truebit,
  author       = {{Rekt News}},
  title        = {Truebit},
  howpublished = {rekt.news},
  year         = {2026},
  url          = {https://rekt.news/truebit-rekt}
}

@misc{rekt_yoprotocol,
  author       = {{Rekt News}},
  title        = {{YO} Protocol's Slippage Bomb},
  howpublished = {rekt.news},
  year         = {2026},
  url          = {https://rekt.news/yo-protocols-slippage-bomb}
}

@article{page2021prisma,
  author  = {Page, Matthew J. and McKenzie, Joanne E. and Bossuyt, Patrick M. and Boutron, Isabelle and Hoffmann, Tammy C. and Mulrow, Cynthia D. and Shamseer, Larissa and Tetzlaff, Jennifer M. and Akl, Elie A. and Brennan, Sue E. and Chou, Roger and Glanville, Julie and Grimshaw, Jeremy M. and Hr\'{o}bjartsson, Asbj\o{}rn and Lalu, Manoj M. and Li, Tianjing and Loder, Elizabeth W. and Mayo-Wilson, Evan and McDonald, Steve and McGuinness, Luke A. and Stewart, Lesley A. and Thomas, James and Tricco, Andrea C. and Welch, Vivian A. and Whiting, Penny and Moher, David},
  title   = {The {PRISMA} 2020 statement: an updated guideline for reporting systematic reviews},
  journal = {BMJ},
  volume  = {372},
  pages   = {n71},
  year    = {2021},
  doi     = {10.1136/bmj.n71}
}

\end{document}